\documentclass[lettersize,journal]{IEEEtran}
\usepackage{amsmath}
\usepackage{graphicx}
\usepackage{caption}
\usepackage{epstopdf}
\usepackage{amsmath,amsfonts,amssymb,amsthm,epsfig,epstopdf,url,array}
\usepackage{graphicx}
\usepackage{color}
\usepackage{textcomp}
\newtheorem{theorem}{Theorem}

\usepackage{cite}
\usepackage{amsthm}
\usepackage{amsmath,amsthm}                                     
\usepackage{marvosym}                                 
\usepackage{array} 
\usepackage{algorithm}
\usepackage{algorithmic}
\usepackage{bm}
\usepackage{hyperref}
\usepackage{subcaption}

\definecolor{yellow}{RGB}{255, 165, 0}

\def\BibTeX{{\rm B\kern-.05em{\sc i\kern-.025em b}\kern-.08em
    T\kern-.1667em\lower.7ex\hbox{E}\kern-.125emX}}
\begin{document}
\title{Semantic HARQ for Intelligent Transportation Systems: Joint Source-Channel Coding-Powered Reliable Retransmissions}
\author{
Yongkang Li,
Xu Wang,
Zheng Shi,
and Yaru Fu
\thanks{Yongkang Li, Xu Wang, and Zheng Shi are with the School of Intelligent Systems Science and Engineering, Jinan University, Zhuhai 519070, China (e-mails: A17200@stu2022.jnu.edu.cn, xuwang@jnu.edu.cn, zhengshi@jnu.edu.cn).}
\thanks{Yaru Fu is with the School of Science and Technology, Hong Kong Metropolitan University, Hong Kong SAR, China (e-mail: yfu@hkmu.edu.hk).}
}
\markboth{Journal of \LaTeX\ Class Files,~Vol.~18, No.~9, September~2020}%
{How to Use the IEEEtran \LaTeX \ Templates}

\maketitle
\begin{abstract}

    The surge of data traffic in Intelligent Transportation Systems (ITS) places a significant challenge on limited wireless resources. Semantic communication, which transmits essential semantics of the raw data, offers a promising solution by reducing redundancy and improving spectrum efficiency.
     However, high vehicle mobility, dynamic channel conditions, and dense vehicular networks severely impact transmission reliability in ITS. To address these limitations, we integrate Hybrid Automatic Repeat reQuest (HARQ) with Joint Source-Channel Coding (JSCC) to provide reliable semantic communications for ITS. To counteract the adverse effects of time-varying fading channels and noise, we propose a generative signal reconstructor module supported by a local knowledge base, which employs a discriminator for channel error detection and a conditional generative network for error correction. We propose three innovative semantic HARQ (sem-HARQ) schemes—Type I sem-HARQ (sem-HARQ-I), sem-HARQ with weighted combining (sem-HARQ-WC), and sem-HARQ with synonymous combining (sem-HARQ-SC)—to enable reliable JSCC-based semantic communications. At the transmitter, both sem-HARQ-I and sem-HARQ-WC retransmit the same semantic signals, while sem-HARQ-SC introduces redundant semantics across different HARQ rounds through synonymous mapping. At the receiver, sem-HARQ-I performs semantic decoding based solely on the currently received signal. In contrast, sem-HARQ-WC enhances reliability by fusing the current received semantic signal with prior erroneous signals at the feature or decision level, thereby exploiting semantic information from failed HARQ rounds. Similarly, sem-HARQ-SC employs feature-level combining, leveraging incremental semantic redundancy to merge semantic features from retransmissions. Experimental results demonstrate that the proposed sem-HARQ significantly improves the reliability of semantic communications compared to a no-HARQ scheme. Among the three schemes, sem-HARQ-I performs the worst due to its inability to utilize failed semantic signals, while sem-HARQ-SC achieves the best performance, albeit with higher implementation complexity.








\end{abstract}
\section{Introduction}



\subsection{Background}
 \IEEEPARstart{T}{he} dramatic expansion of global transportation networks over the past decade has posed critical challenges to urban traffic management. Accelerated urbanization and evolving mobility demands have placed immense pressure on conventional infrastructure, exposing its limitations in managing increasingly complex traffic dynamics. In response, Intelligent Transportation Systems (ITS) have emerged as a pivotal solution, leveraging IoT technologies, real-time data fusion, and intelligent algorithms to enhance traffic efficiency and sustainability \cite{rajkumar2024comprehensive}. However, current ITS architectures remain constrained by traditional communication paradigms that prioritize high data throughput, often struggling to achieve low-latency, context-aware information exchange in vehicle-to-everything (V2X) communications. To overcome this limitation, semantic communication has emerged as a transformative approach. As defined by Shannon, semantic communication represents a higher-level form of communication compared to traditional bit-level transmission \cite{shannon1948mathematical}. In this paradigm, the primary objective shifts from the precise replication of bit sequences to the accurate conveyance of semantic meaning within the transmitted message. This shift enhances resource efficiency, making semantic communication particularly well-suited for bandwidth-constrained vehicular networks \cite{ye2025survey}. Xu \emph{et al.} \cite{10013090} proposed a cooperative semantic-aware architecture for the Internet of Vehicles (IoV) to reduce data traffic by transmitting essential semantics instead of precise symbols, improving spectrum efficiency, and demonstrating its advantages in an image retrieval task within ITS, especially in low signal-to-noise ratio (SNR) environments. Zheng emph{et al.} \cite{10919123} introduced an non-terrestrial network (NTN)-assisted vehicular networks federated learning (NTN-VN-FL) framework to optimize real-time semantic communication codec updates in 6G-era NTN-assisted vehicular networks, using economic models like reverse auctions and Stackelberg games to minimize costs and training delays while maximizing social welfare. Pan \emph{et al.} \cite{10118717} proposed an image segmentation semantic communication (ISSC) system that reduces data transmission while ensuring semantic accuracy. 
 Previous works have applied semantic communication to ITS to improve spectrum efficiency. However, in addition to spectrum efficiency, ITS also requires highly reliable communication \cite{ye2025survey}. The reliability of ITS transmission is easily affected by factors such as high vehicle mobility, dynamic channel conditions, and dense vehicular networks, making it an urgent challenge to ensure semantic reliability in ITS.

\subsection{Related Works}

Joint Source-Channel Coding (JSCC) is an emerging architecture based on deep learning that has attracted considerable research interest in recent years \cite{park2020end}. Unlike conventional separate source-channel coding schemes, JSCC replaces both traditional source coding and channel coding with neural networks that are trained jointly. Notably, Farsad \emph{et al.} \cite{farsad2018deep} demonstrated that the JSCC scheme outperforms traditional separate source-channel coding methods in the regime of finite blocklength. Due to its effectiveness, JSCC has been widely adopted to accommodate semantic communications. For instance, DeepSC \cite{xie2021deep} represents a state-of-the-art implementation of deep learning-enabled semantic communication that introduces a semantic-channel architecture. Specifically, DeepSC employs a Transformer model to perform semantic extraction of text, thereby enhancing the efficiency and accuracy of semantic transmission. Aside from text, semantic communication has also been extended to other modalities. For example, Huang \emph{et al.} in \cite{9953076} proposed a convolutional semantic encoder for extracting semantic information from image data, paired with a Generative Adversarial Network (GAN)-based decoder for reconstruction. Wang \emph{et al.} \cite{9953110} introduced a novel and efficient framework for video semantic transmission, termed DVST, which leverages strong temporal priors between video frames for performance enhancement.

In addition to different communication framework, semantic communication is fundamentally distinguished from conventional communication by its reliance on a dedicated knowledge base. As a critical component of semantic communication, the knowledge base contains the prior information for the transceiver to perform tasks. Especially, knowledge sharing under unstable connectivity among vehicles is a challenge. To address this issues, Wang \emph{et al.} \cite{10847825} introduced a deep semantic communication framework for knowledge sharing (SCKS) in ITS, using generative distillation and GAN-based decoding to enhance bandwidth efficiency and adaptability across different model architectures. 
In addition, different approaches to constructing the knowledge base in semantic communication have been studied. 
Yi \emph{et al.} \cite{10318078} proposed a text-based semantic communication system utilizing shared knowledge bases and introduced an algorithm for constructing such knowledge bases based on similarity comparison methods. Furthermore, Li \emph{et al.} \cite{9928407} developed a domain knowledge-driven semantic communication system, which significantly enhances image recovery performance. In \cite{9953099}, a knowledge base with differentiated access levels is explored, where the receiver has full access to the knowledge, while the transmitter remains unaware of the specific task. This hierarchical access design can enhance the security of semantic communications by providing additional layers of protection. Additionally, Zhou \emph{et al.} \cite{10262128} exploits knowledge graph to construct a knowledge base, which enables semantic communication systems to perform explicit reasoning with improved interpretability.

In the context of semantic communication systems, while significant advancements have been made, the research on ensuring reliability remains in its early stages. Reliability is a critical performance metric, especially emphasized in the development of 5G and beyond networks. Semantic communications face unique challenges in achieving reliability due to the inherent lack of interpretability of neural networks, which are frequently utilized in these systems. The reliability of semantic communications is shaped by multiple factors, with the performance, robustness, and generalization of the encoder and decoder playing crucial roles. These attributes are inherently influenced by the network architecture, training strategies, and dataset quality, collectively determining the system effectiveness. Efforts to enhance reliability include the robust semantic communication system proposed by Peng \emph{et al.} \cite{10486856}, which incorporated a semantic corrector to address and mitigate impairments to the source data. To address the challenge of generalization,  transfer learning was employed to adapt to time-varying channel characteristics among different users \cite{9885016}. Additionally, Chen \emph{et al.} enhanced the generalization of semantic communication systems across diverse tasks through the application of meta-learning techniques \cite{10592533}. Conversely, reliability is compromised by unavoidable physical channel impairments, such as imperfect channel estimation and multipath propagation.
For example, Kim \emph{et al.} \cite{10575676} proposed a channel estimation scheme specifically designed for semantic communication, achieving enhanced estimation accuracy. Similarly, \cite{9252948} utilizes channel state information (CSI) to assist in training, effectively mitigating the adverse impact of fading channels on transmission performance.

Among various techniques for enhancing communication reliability, Hybrid Automatic Repeat reQuest (HARQ) has been widely adopted in modern communication systems. The essence of HARQ lies in leveraging time diversity to improve reliability. Recently, its application in semantic communications has gained growing attention due to its potential to address the unique reliability challenges of semantic-based systems. Specifically, Jiang \emph{et al.} \cite{jiang2022deep} integrated HARQ with Reed-Solomon (RS) channel coding to mitigate transmission errors and improve system robustness. Meanwhile, Zhou \emph{et al.} \cite{zhou2022adaptive} proposed an advanced semantic encoding approach that incorporates multi-bit length selection to optimize HARQ with incremental knowledge, allowing better adaptation to dynamic channel conditions and enhancing overall system reliability and efficiency. However, it is important to note that these HARQ mechanisms rely on conventional channel coding and focus on error detection at the bit level. In contrast, JSCC-enabled semantic communication systems bypass traditional channel coding, presenting a significant challenge in ensuring reliable transmission. To address this limitation, it becomes urgently pivotal to mitigate the effects of physical channel impairments on semantic transmission. Motivated by this challenge, our work explores the integration of HARQ techniques into JSCC-based semantic communication systems to enhance their reliability and robustness.

The HARQ technique is a fundamental method for adapting to challenging channel conditions and enhancing communication reliability. Its integration is crucial for achieving robust semantic communication, as it mitigates errors through an iterative process of error detection, correction, and retransmission. In semantic communication systems that integrate semantic coding with conventional channel coding, traditional HARQ techniques can be directly applied \cite{jiang2022deep, zhou2022adaptive}. However, in JSCC-based architectures, the design and implementation of semantic HARQ (sem-HARQ) remain largely unexplored, necessitating further research to bridge this gap. A major challenge in sem-HARQ lies in the poorly defined notion of semantic error, which complicates the development of mechanisms for detecting, correcting, and retransmitting semantically incorrect information, essential for ensuring reliable communication. Wheeler \emph{et al.} \cite{wheeler2022semantic} proposed a definition of semantic errors based on conceptual space theory, where an error occurs when the receiver reconstructs a conceptual object outside its intended conceptual boundaries. However, this definition is inherently tied to specific conceptual space constructions, limiting its general applicability. Several studies have explored HARQ-inspired mechanisms for semantic communication. Liu \emph{et al.} \cite{10757527} introduced a Double Deep Q-Network (DDQN)-based approach, where retransmission decisions are optimized through performance estimation at the receiver, improving efficiency and reliability. Sheng \emph{et al.} \cite{10734724} proposed an error detection mechanism inspired by HARQ principles for collaborative perception in vehicle-to-vehicle (V2V) communication, demonstrating superior perceptual accuracy and throughput compared to conventional channel coding with separate source coding. Jiang \emph{et al.} \cite{9955991} developed an incremental redundancy HARQ framework (SVC-HARQ) to mitigate transmission errors at key semantic points in video conferencing. For large-scale satellite networks, Gao \emph{et al.} \cite{10695775} introduced a Semantic-Aware Coding and Routing (SACR) mechanism, further refining it into a routing-aware Semantic Adaptive Coding HARQ (SAC-HARQ) strategy, which significantly reduces average transmission delay while increasing effective throughput compared to conventional routing techniques.



However, the aforementioned studies did not incorporate semantic error correction within the HARQ system architecture, nor did they offer a comprehensive analysis of semantic errors warranting retransmission in sem-HARQ. Furthermore, existing approaches lack diverse sem-HARQ schemes tailored to different communication scenarios or adaptable to specific transmission conditions. These limitations hinder the practical deployment and scalability of sem-HARQ in real-world communication systems, underscoring the need for more flexible and adaptive solutions.

\subsection{Contributions}
In this paper, we present a comprehensive semantic HARQ framework for ITS to enhancing the reliability of JSCC-enabled semantic communications by integrating HARQ techniques with advanced generative methods. The proposed model addresses the challenges posed by time-varying fading channels and noise, ensuring robust and efficient data transmission. The main contributions of this work can be summarized as follows.

\begin{itemize}
    \item We present a generative signal reconstructor module that leverages conditional generative network techniques to mitigate adverse effects such as channel fading and noise. This module is designed to correct errors in the received semantic signal before requiring retransmission, marking a significant advancement in receive-side semantic correction within the semantic HARQ framework.
    
    \item We propose various semantic HARQ schemes, including Type I sem-HARQ (sem-HARQ-I), sem-HARQ with weighted combining (sem-HARQ-WC), and sem-HARQ with synonymous combining (sem-HARQ-SC), that consider different methods of semantic information generation and combination. These schemes are tailored to address semantic errors effectively, ensuring data integrity across varying channel conditions.

    At the transmitter, both sem-HARQ-I and sem-HARQ-WC deliver the same semantic signals in retransmissions, while sem-HARQ-SC introduces redundant semantics across different HARQ rounds through synonymous mapping. At the receiver, sem-HARQ-I performs semantic decoding solely based on the currently received signal while sem-HARQ-WC and sem-HARQ-SC fuse the current received semantic signal with prior erroneous signals.
    
    \item A semantic error detector is introduced to generate positive and negative feedback messages by utilizing a transformer-based binary classifier. The detector informs the transmitter of the necessity for retransmission, thereby optimizing the semantic communication process and enhancing reliability.

\end{itemize}
\subsection{Structure of the Paper}
The rest of this paper is outlined as follows. In Section \ref{sec:system model}, we introduce the system model and performance metrics for semantic HARQ communications. Section \ref{sec:HARQ schems} describes the proposed semantic HARQ schemes. The network of sem-HARQ is designed in Section \ref{sec: network}. Section \ref{sec:results} presents simulation results for assessing the proposed sem-HARQ schemes. Section \ref{sec:con} concludes this paper.

\section{System Model and Performance Metrics}\label{sec:system model}
In ITS, ensuring reliable communication is crucial for safety-critical applications, where high vehicle mobility, dynamic channel conditions, and dense network environments pose significant challenges. To address this issue, we propose a novel communication framework that seamlessly integrates HARQ functionality into JSCC-based semantic communication systems. This section first provides an overview of the proposed sem-HARQ transmission process, followed by a description of the architecture of the proposed sem-HARQ system supporting JSCC. A comprehensive discussion of the performance metrics used to evaluate the reliability of the sem-HARQ system is then presented.

\subsection{Sem-HARQ Transmission}
\begin{figure}[htbp]
    \centering
    \includegraphics[width=0.45\textwidth]{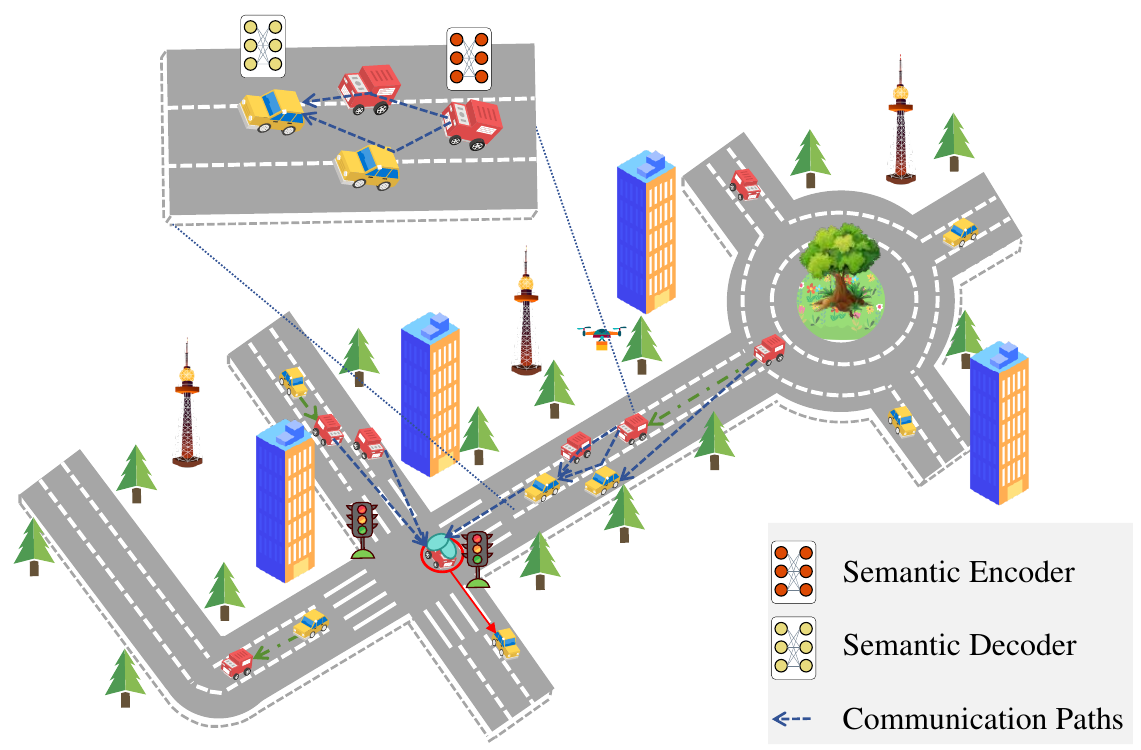} 
    \caption{The framework of proposed sem-HARQ communication system.}
    \label{ITS System Model}
\end{figure}
The ITS system is shown in Fig. \ref{ITS System Model}. The communication process of the proposed sem-HARQ system proceeds as follows. In the case of text transmission, a batch of sentences, denoted as ${\mathbf{S}}_m$, need to be transmitted reliably in the $m$-th HARQ round. The objective of the semantic encoder $f_{\bm{\alpha}}({\bf \cdot})$ is to extract the semantic content from the sentences ${\mathbf{S}}_m$. In the $m$-th HARQ round, the semantic ${\mathbf{x}}_m$ can be denoted as
\begin{equation}
{\mathbf{x}}_m = f_{\bm{\alpha}}({\bf{S}}_m), 
\end{equation}
where $f_{\bm{\alpha}}({\bf \cdot})$ is the semantic encoder and $\bm{\alpha}$ represents the parameter set. As shown in Fig. \ref{ITS System Model}, semantics are transmitted through high-noise wireless channels with complex time-varying conditions, which introduce distortions such as multipath effects, frequency-selective fading, phase noise, and Doppler shifts. After passing through a fading channel, the received semantic ${\mathbf{y}}_m$ in the $m$-th HARQ round can be expressed as
\begin{equation}\label{channel model}
    {\mathbf{y}}_m = \sum_{r=0}^{R-1}{{h_{r,m}}}(t){\mathbf{x}}_m(t-r) + {\mathbf{w}}_m(t),
\end{equation}
where ${{h_{r,m}}}(t)$ is the time-varying gain of the $r$-th path at time $t$, $R$ represents the number of paths, ${\bf{w}}_m(t)$ stands for zero-mean additive white Gaussian noise (AWGN) with variance $\mathcal N$.


Upon receiving ${\bf y}_m$, the receiver employs a novel reconstructor network, $R{\bm{\sigma} }(\cdot)$, to reconstruct normal signals from those affected by channel fading. This approach effectively compensates for signal loss caused by fading, where $\bm{\sigma}$ represents the parameter set of the reconstructor network. This reconstructor network consists of a conditional generative network $G_A(\cdot)$ and a discriminative network $D_A(\cdot)$. This module operates as follows
\begin{equation}
{ {\widetilde{\mathbf{y}}}}_m = 
\begin{cases}
{\mathbf{y}}_m, & \text{if } D_A({\mathbf{y}}_m) = 1, \\
G_A({\mathbf{y}}_m), & \text{if } D_A({\mathbf{y}}_m) = 0.
\end{cases}
\end{equation}
With this module, semantic error correction can be performed before retransmission, thereby reducing transmission latency.
Subsequently, the semantic decoder $g_{\bm{\beta}}({\bf \cdot})$ aims to recover the original sentence ${\mathbf{S}}_m$, resulting in ${{\hat{\mathbf{S}}}}_m$, which is expressed as 
\begin{equation}
    {{\hat{\mathbf{S}}}}_m = g_{\bm{\beta}}({{\widetilde{\mathbf{y}}}}_m).
\end{equation}
For each sentence, based on the results of the error detector $D_{\bm{\delta}}({{\hat{\mathbf{S}}}}_m)$, if the error corrector still cannot successfully correct the error, the receiver sends feedback NACK to the transmitter, requesting a retransmission for this sentence in next batch. 
Conversely, if the error detector accepts this decoded sentence, the receiver sends a feedback ACK to the transmitter. This process is repeated until the transmitter receives an ACK or the maximum number of retransmissions, $M$, is reached.

\subsection{System Model}
\begin{figure}[htbp]
    \centering
    \includegraphics[width=0.45\textwidth]{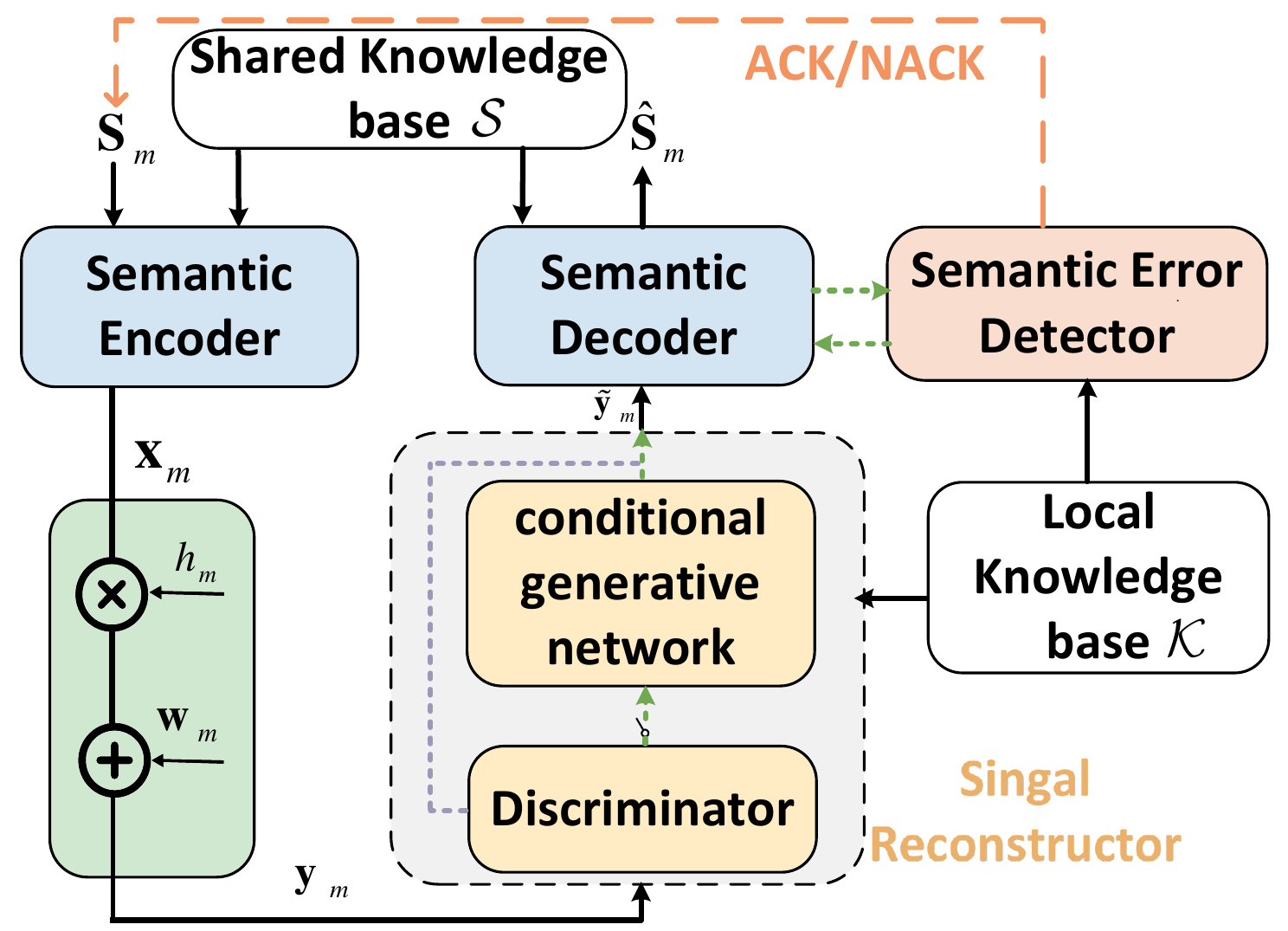} 
    \caption{The framework of proposed sem-HARQ communication system.}
    \label{System Model}
\end{figure}

The overall architecture of the system is illustrated in Fig. \ref{System Model}. The system architecture is divided into three parts: the knowledge base, the semantic communication system, and the HARQ enhancement module.

\subsubsection{Knowledge Base}
Similar to \cite{9953099}, the knowledge base of the proposed Sem-HARQ scheme is designed hierarchically and consists of two components, a shared knowledge base $\mathcal{S}$ and a local knowledge base $\mathcal{K}$. The shared knowledge base, $\mathcal{S}$, is accessible to both the transmitter and receiver and contains knowledge that supports their semantic transmission tasks. In this paper, we focus on text transmission, with the knowledge in $\mathcal{S}$ modeled as a large corpus. By extracting underlying patterns from $\mathcal{S}$, the transmitter gains the ability to extract semantics, while the receiver acquires the capability to recover the text. The local knowledge base, $\mathcal{K}$, contains task-specific knowledge that is exclusively accessible to the receiver. In the proposed sem-HARQ scheme, $\mathcal{K}$ provides the information necessary for the receiver to detect and correct semantic errors. This approach addresses the issue of compatibility between JSCC-enabled architecture and HARQ. The modeling and the acquisition of this local knowledge, which are essential for enabling these functionalities, remain an underexplored area. We will construct a relevant local knowledge base tailored for specific fading channels and provide a detailed discussion and analysis in Section \ref{sec: network}.

\subsubsection{Semantic Communication System}
The semantic communication system comprises a semantic encoder and a semantic decoder, denoted as $f_{\bm{\alpha}}({\bf \cdot})$ and $g_{\bm{\beta}}({\bf \cdot})$, respectively, where $\bm{\alpha}$ and $\bm{\beta}$ represent the parameter sets of each component.
These functions are implemented as deep neural networks, responsible for extracting semantics from the transmitted message and reconstructing the message from the received semantic information, respectively. Furthermore, both functions are jointly trained in an end-to-end manner using a shared knowledge base $\mathcal{S}$. 
As the core module of the proposed system, the semantic communication system must adapt its network selection based on the modality of the transmitted information. To illustrate, consider a text transmission system: the encoder $f_{\bm{\alpha}}({\bf \cdot})$ consists of multiple Transformer encoder layers followed by several dense layers. The Transformer encoder layers extract semantic features from the input text, while the dense layers adapt the channel for transmission. Conversely, $g_{\bm{\beta}}({\bf \cdot})$ comprises several dense layers followed by Transformer decoder layers. The dense layers restore the semantic representation, while the Transformer decoder layers reconstruct the original text.

\subsubsection{Reconstruction and Detector Modules for HARQ}
To achieve HARQ's dual objectives of error correction prior to retransmission and feedback generation, the proposed HARQ modules incorporates two deep neural network-based components at the receiver: a reconstructor $R_{\bm{\sigma}}({\bf \cdot})$ and a semantic error detector $D_{\bm{\delta}}({\bf \cdot})$, both trained independently. Here, $\bm{\delta}$, and $\bm{\sigma}$ denote the parameter sets associated with each component.
The reconstructor employs a conditional generative network $G_A(\cdot)$ to dynamically regenerate semantically corrected signals by leveraging detected error patterns from corrupted inputs and domain-specific knowledge from the local base $\mathcal{K}$. This process is further refined through a discriminative network $D_A(\cdot)$, which ensures semantic fidelity. Meanwhile, the semantic error detector $D_{\bm{\delta}}({\bf \cdot})$ evaluates the reconstructed text for residual semantic error, generating feedback to the transmitter based on its output.

\subsection{Performance Metrics of Semantic Communications}
 The goal of the proposed sem-HARQ system is to transmit the underlying semantics and reconstruct the original sentences at the receiver based on received semantics, rather than focusing on transmitting symbols with high accuracy. Therefore, traditional metrics such as Bit Error Rate (BER) do not adequately reflect the performance of this system. To evaluate the performance of the proposed system, commonly used metrics in natural language processing, such as  BLEU scores and semantic similarity, are employed.
\subsubsection{BLEU Score}
The BLEU score is a word-level evaluation metric that assesses semantic differences between two sentences by quantifying the word-level discrepancies between the transmitted and recovered sentences. 
The BLEU score between $\mathbf{s}$ and $\hat{\mathbf{s}}$ is defined as
\begin{equation}
    \log {\rm {BLEU}} = 
 [1-{{{l_\mathbf{s}}}}/{l_{\hat{\mathbf{s}}}}]^- + \sum\nolimits_{n = 1}^N {{u_n}\log {f_n}}, 
\end{equation}
where $[x]^-=\min \left\{x,0\right\}$, $l_\mathbf{s}$ and $l_{{\hat{\mathbf{s}}}}$ represent the lengths of the original sentence $\mathbf{s}$ and the recovered sentence ${\hat{\bf{s}}}$, respectively, $u_n$ is the weights of $n$-grams, and $f_n$ indicates the $n$-grams score that can be calculated as
\begin{equation}\label{eqn:bleu_e}
    f_{n}=\frac{\sum_{k} \min \left\{C_{k}({{\hat{\mathbf{s}}}}), C_{k}(\mathbf{s})\right\}}{\sum_{k}  C_{k}({{\hat{\mathbf{s}}}})},
\end{equation}
where $C_{k}(\cdot)$ is the function counting the frequency of the $k$-th element in $n$-grams and $k \in [1,L-n+1]$.
\subsubsection{Sentence Similarity}
Although BLEU scores provide an intuitive metric, they have notable limitations. First, BLEU operates at the word level, disregarding sentence grammar and word order. Second, even when words with similar meanings are recovered, the BLEU score may still be low. To address these issues, a BERT-based sentence similarity metric has been proposed. BERT is a pre-trained, decoder-only large language model containing billions of parameters that encodes input sentences into a semantic space, where sentences with similar meanings are positioned closer together. The sentences $\mathbf{s}$ and $\hat{\mathbf{s}}$ are each encoded by BERT to generate semantic vectors, after which sentence similarity can be computed as follows,
\begin{equation}\label{Simarlity}
    \operatorname{Sim}({\mathbf{s}},\hat{\mathbf{s}}) = \frac{{\operatorname{BERT}}({\bf s}) \cdot {\operatorname{BERT}}(\hat{\mathbf{s}})^T}{||{\operatorname{BERT}}({\mathbf{s}})||\cdot ||{\operatorname{BERT}}(\hat{\mathbf{s}})|| },
\end{equation}
where $\operatorname{BERT}$ denotes the pre-trained large model, which is capable of efficiently extracting sentence-level semantics.

\section{Sem-HARQ Schemes and Semantic Combination Strategy}\label{sec:HARQ schems}
Conventional bit-level HARQ schemes are no longer suitable for the proposed sem-HARQ system, as retransmissions in sem-HARQ are triggered by semantic errors rather than traditional bit errors. To address this, this section introduces three novel sem-HARQ schemes specifically designed for semantic retransmissions within the sem-HARQ framework. The first scheme operates without reusing semantic error information from previous transmissions, while the latter two schemes incorporate prior erroneous semantics to enhance reliability. Moreover, in the first two schemes, the retransmitted semantic content remains unchanged across retransmissions. In contrast, the third scheme progressively introduces additional redundant semantics in each retransmission to further improve reliability.
Additionally, this section explores two distinct methods for utilizing erroneous semantics: semantic fusion and confidence score-weighted combination. These approaches aim to optimize the exploitation of semantic information during retransmissions, thereby enhancing communication robustness.

\subsection{sem-HARQ-I}
\begin{figure}[htbp]
    \centering
    \includegraphics[width=0.45\textwidth]{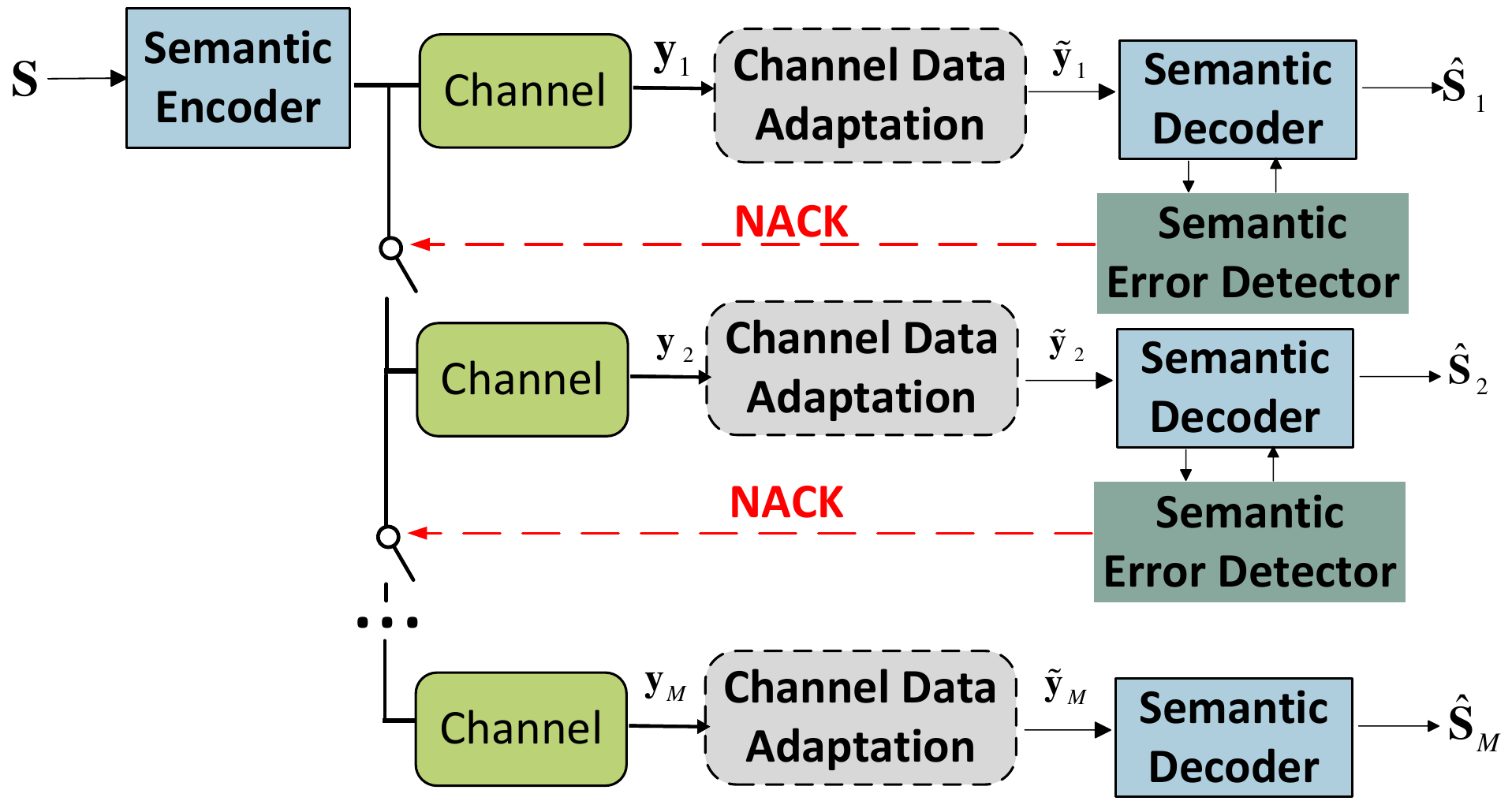} 
    \caption{The proposed sem-HARQ-I scheme.}
    \label{Fig:HARQI}
\end{figure}
The first proposed scheme, referred to as sem-HARQ-I, employs a straightforward retransmission mechanism for semantic information. In each HARQ round, if the receiver identifies semantic errors in the decoded sentence, the erroneous semantic data is discarded, and the transmitter retransmits the identical semantic content. 

As illustrated in Fig. \ref{Fig:HARQI}, the communication process for sem-HARQ-I scheme as follows. $\mathbf{S}$ needs to be reliably transmitted. The transmitter employs a semantic encoder $f(\cdot)$ to extract semantics from $\mathbf{S}$, generating $\mathbf{x} = f(\mathbf{S})$. The semantics $\mathbf{x}$ traverse a fading channel in the $m$-th HARQ round which can be expressed as (\ref{channel model}). 
 At the receiver, ${\mathbf{y}}_m$ is initially analyzed by the discriminant function $D_A(\cdot)$ within the reconstructor module. If ${\mathbf{y}}_m$ is classified as abnormal data, the conditional generative network $G_A(\cdot)$ is applied, yielding $\mathbf{\widetilde{y}}_m = G_A({\mathbf{y}}_m)$. If $D_A(\cdot)$ classifies ${\mathbf{y}}_m$ as normal, no transformation is applied, and ${\mathbf{y}}_m$ is used directly. Subsequently, the received semantics or the transformed data is decoded by the semantic decoder $g(\cdot)$ to obtain the recovered sentence ${\bf{\hat{S}}}$. Then ${\bf{\hat{S}}}$ is discriminated using the semantic error discriminator, obtaining the probability $\hat{p}$ that ${\bf{\hat{S}}}$ belongs to a normal sentence. If $\hat{p} > \lambda$, it is considered that the recovered sentence ${\bf{\hat{S}}}$ is semantically correct. The receiver will then accept ${\bf{\hat{S}}}$ and send an acknowledgment ($ACK$) to the transmitter to indicate successful reception. Otherwise, if ${\bf{\hat{S}}}$ is judged to have semantic errors ($\hat{p} \leq \lambda$), it will be directly discarded, and a negative acknowledgment ($NACK$) will be sent to the transmitter. The transmitter will resend ${\bf{\hat{S}}}$, and the receiver will repeat the above steps, attempting to decode the resent sentence. This process continues until the sentence is successfully received or the maximum retry limit is reached.
Specifically, the first type of HARQ scheme involves direct retransmission. The transmitter does not require additional operations, and erroneous sentences are directly discarded without the need for extra memory.

\subsection{sem-HARQ-WC}
\begin{figure}[t]
    \centering
    \begin{minipage}[b]{0.45\textwidth}
        \centering
        \includegraphics[width=\linewidth]{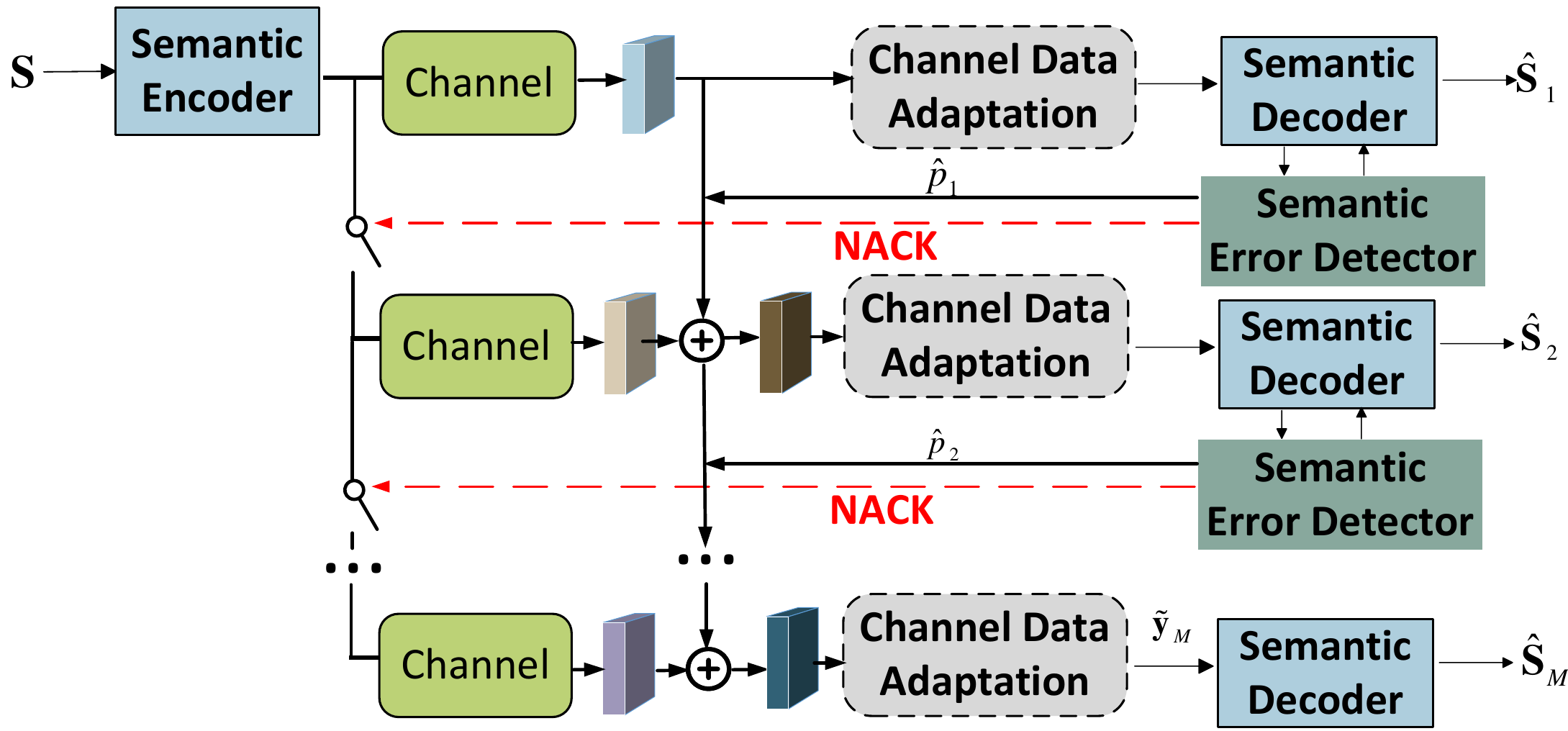}
        \subcaption{sem-HARQ-WC-FC}
        \label{Fig:HARQFC}
    \end{minipage}
    \hfill
    \begin{minipage}[b]{0.45\textwidth}
        \centering
        \includegraphics[width=\linewidth]{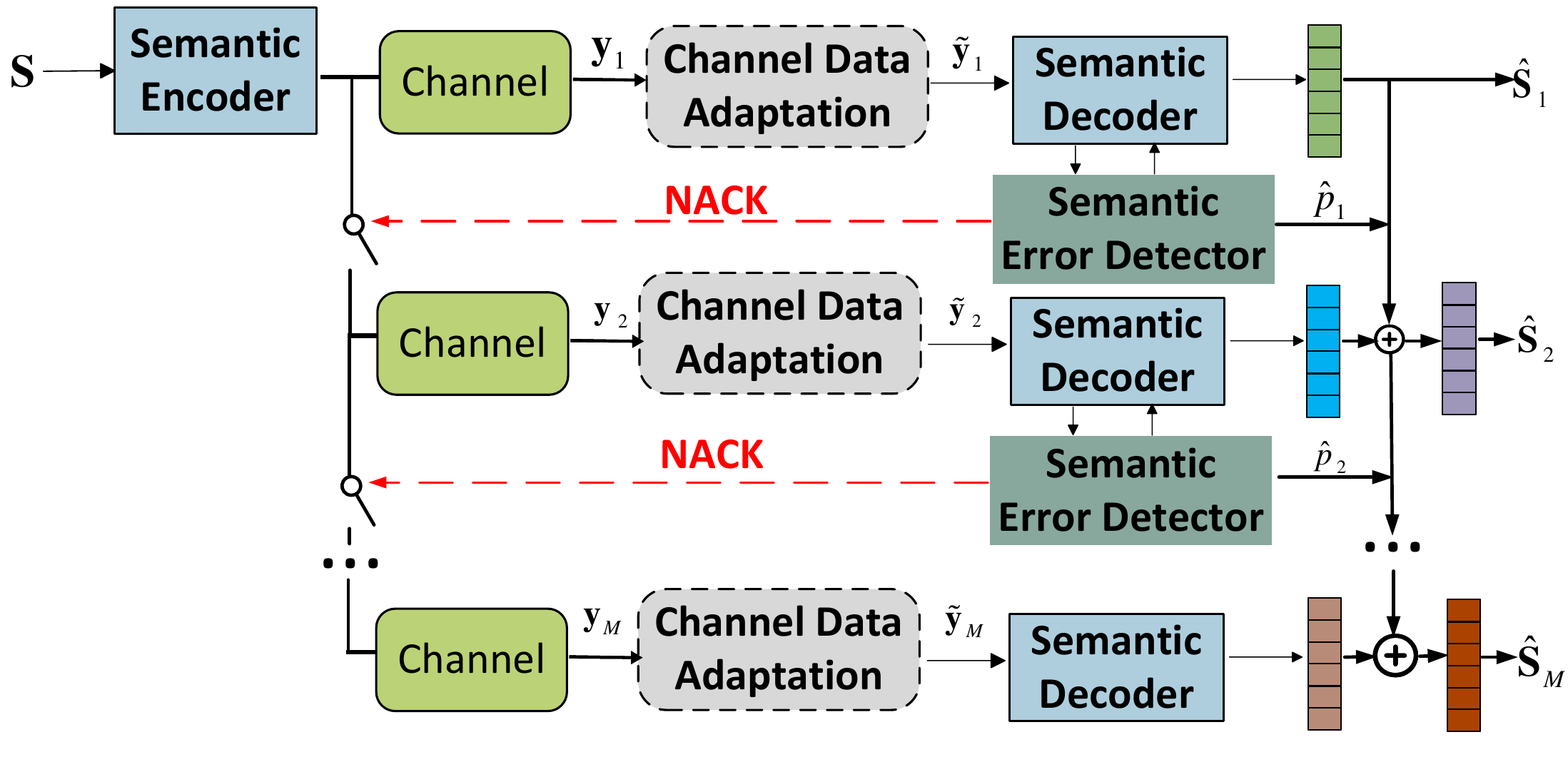}
        \subcaption{sem-HARQ-WC-DC}
        \label{Fig:HARQDC}
    \end{minipage}
    \caption{The proposed sem-HARQ-WC schemes. The first variant, sem-HARQ-WC-FC, involves merging semantic features prior to decoding, whereas the second variant, sem-HARQ-WC-DC, performs semantic decoding first, followed by the combination of decoded word probability results.}
    \label{Fig:HARQ}
\end{figure}

The advantages of the sem-HARQ-I scheme lie in its simplicity and low memory requirements, making it particularly suitable for receivers with limited computational resources. However, its primary limitation is the direct discarding of erroneous messages, which leads to poor utilization of previously received data. This, in turn, results in an increased number of retransmissions and, consequently, higher latency. By contrast, preserving the semantics of previously received messages to assist in subsequent decoding can significantly enhance system performance. However, this approach imposes greater demands on the receiver, requiring higher computational and memory resources. This enhanced scheme, which retains previously received semantics for further decoding attempts, is referred to as sem-HARQ-WC. By leveraging the preserved semantic information from earlier transmissions, the receiver can better reconstruct the original message in subsequent rounds, improving communication efficiency and reducing retransmission requirements. Similar to sem-HARQ-I, sem-HARQ-WC deliver the same semantic signals in retransmissions. The key difference lies in the receiver's handling of semantic errors. When a semantic error is detected in the received message, the received semantic features is preserved for subsequent processing.

In sem-HARQ-WC, a semantic weighted combination approach is proposed to fuses the current received semantic signal with prior erroneous signals.
In the first $m$ rounds of HARQ, the received semantic are denoted as $\mathbf{B}_1, \mathbf{B}_2, \dots, \mathbf{B}_m$. The combined output can be expressed as
\begin{equation}
    {\mathbf{B}}_{m}^{\text{cb}} = \sum_{i=1}^m {a_i \mathbf{B}_i}, \quad \text{where } a_i = \frac{\hat{p}_i }{\sum_{k=1}^m \hat{p}_k}.
\end{equation}
where $a_i$ represents the weighting factor for each received semantic. In other words, each round of HARQ is independently decoded to obtain the received confidence scores prior to semantic combination. The higher the confidence score, the greater the weighting coefficient, and consequently, the larger the contribution to the combined semantics.
For comparison, a simple equal combination method is proposed that assigns equal weighting coefficients to each HARQ round, in other words,  $a_i = \frac{1}{m}$ for any $i\in [m]$.

Nevertheless, determining which specific information should be preserved remain open research questions. To address this, this subsection introduces two distinct sem-HARQ-WC schemes: sem-HARQ-WC-FC, which retains feature-level information, and sem-HARQ-WC-DC, which preserves decision-level information.
\subsubsection{sem-HARQ-WC-FC}
In sem-HARQ-WC-FC, the received semantic features are fused at the feature level to form a combined representation. This fused representation is then used as the input to the semantic decoder, producing the final combined decoded sentence.
The communication process of sem-HARQ-WC-FC is illustrated in Fig. \ref{Fig:HARQFC}. 
After passing through the fading channel, the receiver obtains the semantic feature $\mathbf{y}_m$. 
Subsequently, the conditional generative network generate the corrected semantic features, which can be expressed as
\begin{equation}
\mathbf{\widetilde{y}}_m = 
\begin{cases} 
\mathbf{y}_m, & \text{if } D_A(\mathbf{y}_m) = 1, \\ 
G_A(\mathbf{y}_m), & \text{if } D_A(\mathbf{y}_m) = 0.
\end{cases}
\end{equation}
The received semantic feature $\mathbf{\widetilde{y}}_m$ is merged at this stage to produce the combined semantic features ${\mathbf{y}}_{m}^{\text{FC}}$.
Subsequently, the combined semantic representation ${\mathbf{y}}_{m}^{\text{cb}}$ is decoded by the semantic decoder $g_{\beta}(\cdot)$ to generate the combined decoded sentence ${\mathbf{S}}_{m}^{\text{cb}}$. Feedback is generated in the same way as previously mentioned.

\subsubsection{sem-HARQ-WC-DC}
To enhance interpretability, this subsection introduces the sem-HARQ-WC-DC scheme, which differs from sem-HARQ-WC-FC in that semantic combining occurs at the decision level rather than at the feature level. The communication process of sem-HARQ-WC-DC is illustrated in Fig.\ref{Fig:HARQDC}. Overall, the communication process remains the same as sem-HARQ-WC-FC, with the distinction being the stage at which the merging operation is performed.
At the decision level, in the $m$-th HARQ round, the decoder first processes the received semantics to generate word probability vectors ${\mathbf{\mathcal{D}}}_m$.  A greedy algorithm is then applied to each probability vector, selecting the word with the highest probability iteratively until the entire sentence is reconstructed. Unlike sem-HARQ-WC-FC, which merges semantic features before decoding, sem-HARQ-WC-DC retains and combines word probability vectors across multiple HARQ rounds.

\subsection{sem-HARQ-SC}
\begin{figure}[htbp]
    \centering
    \includegraphics[width=0.45\textwidth]{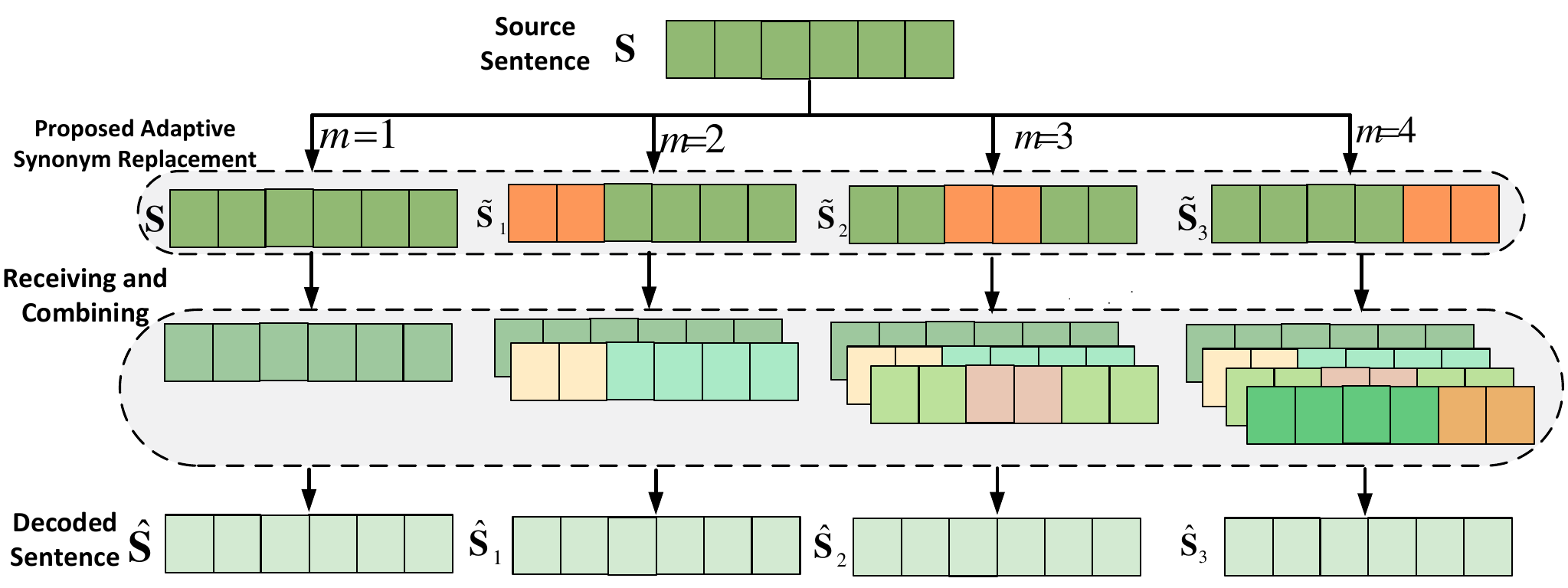} 
    \caption{The proposed adaptive synonym replacement method.}
    \label{IS_Combine}
\end{figure}
\begin{figure}[htbp]
    \centering
    \includegraphics[width=0.45\textwidth]{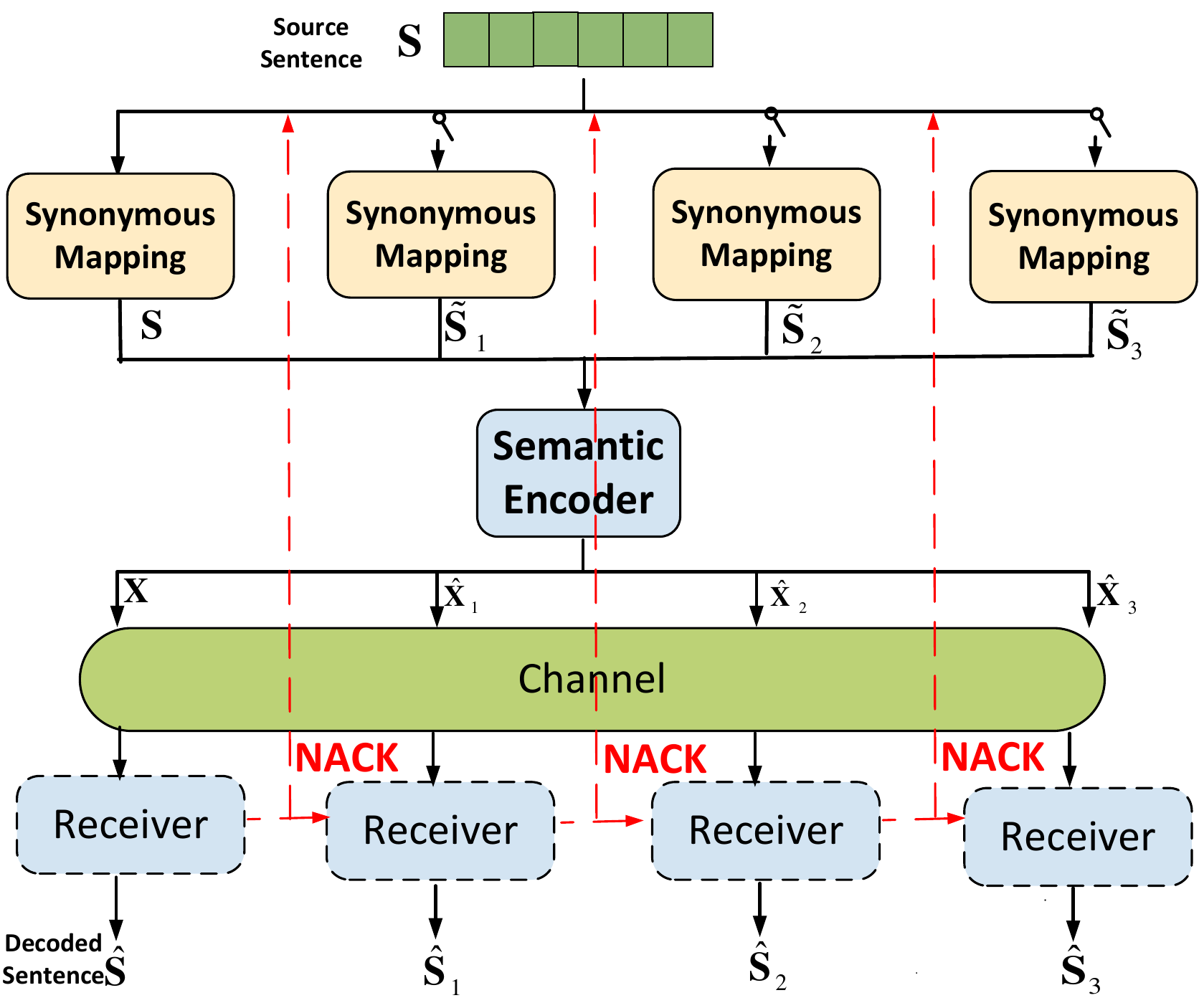} 
    \caption{The proposed sem-HARQ-SC scheme.}
    \label{HARQ_IS}
\end{figure}
This subsection introduces the third proposed scheme, sem-HARQ-SC, which differs from sem-HARQ-WC by incorporating a synonymous mapping step before each retransmission, followed by the extraction of redundancy for retransmission. The principle underlying this scheme is rooted in the synonymous and polysemy characteristics of natural language. In semantic communication, both polysemy and synonymous are essential properties that cannot be overlooked. Polysemy refers to the phenomenon where a word may have multiple meanings, potentially leading to semantic confusion. As a result, contextual understanding is required to interpret the intended meaning correctly. When a word exhibits higher polysemy, indicating it is associated with a broader and more diverse set of meanings within the semantic space, it becomes increasingly prone to being interpreted with incorrect or unintended meanings due to signal degradation during transmission. Synonymous, on the other hand, implies that a single semantic meaning can have multiple equivalent expressions. This characteristic forms the foundation of semantic communication, where the transmitter needs only to convey the semantic meaning, and the receiver can select any expression from a set of equivalent expressions with the same semantic meaning. In human communication, when the receiver struggles to understand the intended meaning, the sender often restates the expression using less ambiguous phrasing to facilitate comprehension. This process involves selecting an expression with reduced polysemy from a set of synonyms, enabling the receiver to integrate the new expression with the previous context for a better understanding. The principle of sem-HARQ-SC follows a similar approach. When the receiver detects a semantic error, the transmitter performs a synonymous mapping to extract alternative semantic representations, which are then retransmitted as redundancy. The receiver subsequently reconstructs the original sentence by integrating the retransmitted redundancy with the previously received semantic. The introduction of synonymous redundancy provides additional information to the receiver, thereby enhancing the accuracy of semantic interpretation and reducing the likelihood of semantic errors. 

The implementation of sem-HARQ-SC requires synonymous mapping algorithms. 
In this paper, we propose an adaptive synonym replacement method to achieve the desired synonymous mapping. 
Specifically, let the sentence to be transmitted be $\mathbf{S}= [w_1, w_2, \dots , w_i, \dots , w_L]$, where $w_i$ denotes the $i$-th word in sentence $\mathbf{S}$, $L$ is the length of the sentence. The core principle of the synonym replacement algorithm is to select synonyms from a synonym set and replace the words in a sentence with their corresponding synonyms.
Let $\mathbf{W}_i$ represent the set of synonyms for $w_i$, which is defined as $\mathbf{W}_i = [w_i, w_i^1, \dots , w_i^j, \dots , w_i^J]$, where $w_i^j$ denotes the $j$-th synonym of $w_i$. The size of the synonym set is $J+1$, which is a non-empty set, meaning that $\mathbf{W}_i \neq \emptyset$. 
As illustrated in Fig. \ref{IS_Combine}, during the $m$-th retransmission, the words from $L(m-1)/M$ to $Lm/M$ are substituted with their synonyms. This process can be expressed as $\mathbf{S}_m = \psi (\mathbf{S})$, where $\mathbf{S}_m = [w_1, \dots, w_{L(m-1)/M}^m, \dots, w_{Lm/M}^m, \dots, w_L]$.
The receiver subsequently combines the semantics from multiple transmissions along with their synonymous redundancies, as shown in Fig. \ref{IS_Combine}. 
The substituted synonyms are carefully chosen and positioned within the sentence, which ensures that each word gains synonymous redundancy in the final combined semantics. This approach helps enhance the robustness of the semantic representation in the context of sentence paraphrasing.

The communication scheme of sem-HARQ-SC is shown in Fig. \ref{HARQ_IS}. The initial transmission of $\mathbf{x}$ does not include synonymous semantics. The receiver then performs semantic error detection on the received data and returns a NACK if an error is detected or an ACK if the data is correctly decoded.  Upon receiving the NACK, the transmitter applies a synonymous mapping to the $T(\cdot)$ and retransmits the synonymously mapped semantics corresponding to $\mathbf{\widetilde{S}}_m$. The receiver subsequently combines $\mathbf{y}^T_m$ with the previously received semantics for semantic decoding. If the error persists, retransmissions continue until semantic decoding is successful or the maximum allowed retransmissions, $M$, are reached.

\section{Reconstruction and Detector Modules for HARQ}\label{sec: network}
This section presents the design and architecture of the HARQ modules in the proposed system. We first discuss the concept of semantic errors and their impact on communication reliability. Then, we introduce a novel generative signal reconstruction module and a semantic error detector. Finally, the section details the generation of the local knowledge base and outlines the training process of the HARQ enhancement module.

\subsection{Problem Definition}


Implementing semantic HARQ requires addressing two critical challenges: identifying the specific types of semantic errors that should prompt retransmissions, and devising efficient methods to accurately detect these errors.

Existing research, such as the analysis using Conceptual Spaces \cite{9991159}, provides valuable insights but is limited by its focus on the color domain, making generalization to other contexts challenging.
For text transmission, consider a sentence ``weather is good today" that needs to be transmitted.
In the current communication system, $\mathbf{x}_{bit}$ is the signal to be sent, which is transmitted over the fading channel.
The receiver then receives $\mathbf{y}_{bit}$.
Suppose the decoded sentence is ``today's weather is good"
As the bit-level information differs, i.e., $\mathbf{x}_{bit} \neq \mathbf{y}_{bit}$, the error detection module typically identifies it as an error and sends a NACK to the transmitter, triggering a retransmission. This is referred to as a bit error. While this approach ensures transmission reliability, it does so at the cost of increased latency. However, if we consider the transmitted and received sentences rather than individual bits, we may find that they still convey the same meaning despite bit-level discrepancies. In other words, from a semantic perspective, the transmission remains successful. This observation answers the first key question: In semantic communication, the primary objective is to ensure the reliable transmission of semantics. Retransmission should be triggered only when a semantic error occurs, rather than merely upon detecting a bit error.

We now turn to the second issue, which involves determining whether the received sentence contains semantic errors.
It is clear that minor inaccuracies in a sentence may not hinder the comprehension of its meaning. In other words, there exists a tolerable distortion threshold $\Theta$. When the distortion surpasses this threshold, it can be concluded that the received sentence contains semantic errors. Semantic distortion is quantified by the function $d(\mathbf{S},\hat{\mathbf{S}})$, and when the distortion exceeds the threshold, i.e., $d(\mathbf{S},\hat{\mathbf{S}}) \geq \Theta$, a semantic error is deemed to have occurred. Let the data distribution of the decoded sentences be represented by $P_{\hat{\mathbf{S}}}$. It can be shown that the received sentence can be classified into two distinct semantic distribution spaces, $P_{\mathbf{S}_1}$ is the set of all decoded sentences $\hat{\mathbf{S}} = g(H(f(\mathbf{S})))$ that satisfy $d(\mathbf{S}_i,\hat{\mathbf{S}}_i) \leq \Theta$, while $P_{\mathbf{S}_2}$ is the set of all decoded sentences $\hat{\mathbf{S}} = g(H(f(\mathbf{S})))$ that satisfy $d(\mathbf{S}_i,\hat{\mathbf{S}}_i) > \Theta$. It is easy to observe that $P_{\hat{\mathbf{S}}} = P_{\mathbf{S}_1} \cup P_{\mathbf{S}_2}$ and $P_{\mathbf{S}_1} \cap P_{\mathbf{S}_2} = \emptyset$. This demonstrates that the received sentences can be divided into semantically valid and semantically erroneous categories. As a result, the problem of identifying semantic errors in the received sentence can be framed as a classification task. In practice, however, the threshold $\Theta$ is often not well-defined, and the measurement of semantic distortion lacks a clear definition as well. Due to this factor, the labels in the training samples may not correspond to their true values. This necessitates the characterization of the general behavior of the semantic distortion generalization gap, which quantifies the difference between the semantic distortion generalization error and the semantic distortion empirical error.

\subsection{Generative Signal Reconstruction Module}

 \begin{figure}[htbp]
    \centering
    \includegraphics[width=0.45\textwidth]{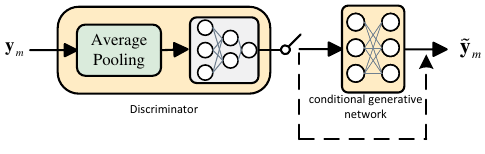} 
    \caption{The reconstructor architecture.}
    \label{Semantic Reconstructor model}
\end{figure}
The proposed reconstructor architecture is shown in Fig. \ref{Semantic Reconstructor model}. It is composed of a conditional generative network $G_A(\cdot)$ and a discriminative network $D_A(\cdot)$. The generative network $G_A(\cdot)$ reconstructs a recovered signal that closely approximates the original signal based on the corrupted input, while the discriminative network $D_A(\cdot)$ determines whether a given semantic signal contains errors. The semantic signal that has been identified as erroneous by $D_A(\cdot)$ will be sent as input to $G_A(\cdot)$ as the conditions to guide the signal reconstruction.

$G_A(\cdot)$ is a fully connected neural network consisting of an input layer, three hidden layers and an output layer. The input of $G_A(\cdot)$ is the semantic signal that needs to be reconstructed, denoted as $\mathbf{{\widetilde{y}}} = G_A(\mathbf{x},\mathbf{H})$, where $\mathbf{H}$ is the condition. To effectively model the complex relationship between the abnormal and correct distributions, the size of the three hidden layers of $G_A(\cdot)$ are $4\mathcal{V}$, $16\mathcal{V}$ and $4\mathcal{V}$, respectively, where $\mathcal{V}$ represents the input dimension and the output dimension. The final output of the model is the generated semantic signal. The discriminative network $D_A(\cdot)$ takes as input the average-pooled received semantic signal and consists of one input layer, three hidden layers, and one output layer. The size of the three hidden layers of $D_A(\cdot)$ are $\frac{\mathcal{V}}{2}$, $\frac{\mathcal{V}}{4}$ and $\frac{\mathcal{V}}{8}$. The output of $D_A(\cdot)$ is the discriminative result, indicating whether the semantic signal is erroneous. To mitigate the vanishing gradient problem while maintaining nonlinearity, all neurons in the hidden and output layers adopt the Rectified Linear Unit (ReLU) activation function.

\subsection{Semantic Error Detector}


 \begin{figure}[htbp]
    \centering
    \includegraphics[width=0.45\textwidth]{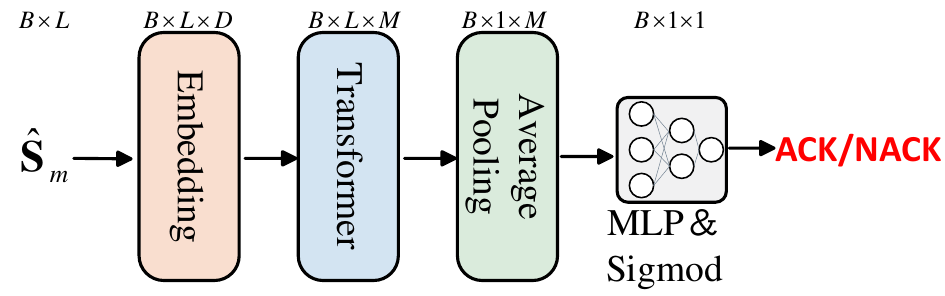} 
    \caption{Semantic Error detector model.}
    \label{Semantic error detector model}
\end{figure}

The proposed semantic error detector $D_{\bm{\delta}}({\bf \cdot})$ is depicted in Fig. \ref{Semantic error detector model}. 
The network takes the received sentences $\bf \hat{S}$ as input and outputs the confidence level for accepting the sentence. The semantic error detector comprises an embedding layer, a multi-head Transformer, and a multilayer perceptron. The embedding layer represents the input sentence into mathematical space $\mathbb{W}$.  Through the embedding layer, the sentences can be represented as a dense word vector $\mathbf{D} \in \mathbb{R}^{B \times L \times D}$, where $B$ is the batch size, $L$ is the sentence length and $D$ is the word vector dimension size.
The multi-head Transformer consists of three Transformer layers, which helps the model capture a richer diversity of relationships between inputs. A Transformer layer consists of a self-attention layer and a feed-forward layer.
The self-attention layer uses the self-attention mechanism to help the model understand the relationship between words. In this layer, each word will observe the surrounding words to find out the contextual information related to itself. The feed-forward layer consists of two linear layers to further process the information obtained by the self-attention layer. In this layer, each word will integrate and process information based on the context information it has collected. Pass the Transformer, the semantic information $\mathbf{M} \in \mathbb{R}^{B \times L \times M}$ conveyed by $\bf \hat{S}$ is obtained, where $M$ is the output dimension of Transformer.
Sentence embeddings $\mathbf{P} \in \mathbb{R}^{B \times 1 \times M}$ are subsequently derived by applying average pooling to the word embeddings. 
The multilayer perceptron consists of one input layer with two hidden layers and an output layer. The input layer receives the sentence embedding, and then the sentence embedding is fed into the hidden layer to get the confidence score $\hat{p}$ of each sentence, indicating the probability that the sentence is semantically correct. 

\subsection{Local Knowledge Base Generation and Training}
In order for the proposed HARQ enhancement modules to acquire the capabilities of semantic signal reconstruction and semantic error detection, it is necessary to establish a local knowledge base for specific fading channels. To achieve this, a trained encoder-decoder is used to perform multiple simulated transmissions under the given fading channel while collecting relevant information.
For the transmitted sentence pair$(\mathbf{S},\hat{\mathbf{S}})$, considering that minor word errors or omissions generally do not hinder sentence comprehension and align with the selected performance metric, we classify sentences with BLEU scores above 0.9 as belonging to the normal observation space, while those with scores below this threshold are categorized as part of the abnormal observation space. Specifically, when the BLEU score is below 0.9, the classification label $C$ is set to 0; otherwise, it is set to 1.
For $D_A(\cdot)$, the sample is constructed as $ ({\mathbf{\hat{y}}}, C)$. Additionally, when $C=0$, the semantic representation ${\mathbf{y}}$ extracted by the transmitter is retained to form the sample $({\mathbf{y}}, {\mathbf{\hat{y}}})$ for training $G_A(\cdot)$. For $D_{\bm{\delta}}({\bf \cdot})$, the required sample is constructed as $({\bf\hat{S}}, C)$. Once sufficient observational data is collected, the three components undergo independent supervised training. Hence, The constructed local knowledge base can be represented as $\mathcal{K} =(\mathcal{K}_1,\mathcal{K}_2,\mathcal{K}_3)$, where $\mathcal{K}_1 =\left\{({\mathbf{\hat{y}}}_1, C_1),..., ({\mathbf{\hat{y}}}_n, C_n),\right\}$ $\mathcal{K}_2 = \left\{ ({\mathbf{y}}_{n+1}, {\mathbf{\hat{y}}}_{n+1}),...,({\mathbf{y}}_{n+m}, {\mathbf{\hat{y}}}_{n+m})\right\}$ and $\mathcal{K}_3 = \left\{
({\bf \hat{S}}_{n+m+1}, C_{n+m+1}), ..., ({\bf \hat{S}}_{n+m+k}, C_{n+m+k})\right\}$.

To train $D_A(\cdot)$, a mini batch of samples is selected from $\mathcal{K}_1$. It contains randomly selected positive and negative examples.
The semantic signal is entered into $D_A(\cdot)$ to obtain the correct probability $p$, which can be expressed as
\begin{equation}
    p = D_A(\mathbf{\hat{y}}).
\end{equation}
To make $p$ as close as possible to its label, binary cross-entropy (BCE) is used as the loss function, which can be expressed as 
 \begin{equation}\label{loss}
     \mathcal{L}_{\text{BCE}} = -\frac{1}{N}\sum_{i=1}^{N}\bigg(C_i\cdot\log{p_i} + (1-C_i)\cdot \log(1- p_i)\bigg),
 \end{equation}
where $N$ is the sample size, $p_i$ is the predicted probability for $i$-th sample. The Adam optimization algorithm is used to iteratively update the parameters during the training.
To train $G_A(\cdot)$, a mini batch of samples is adopted from $\mathcal{K}_1$. The degraded semantic signal $\mathbf{\hat{y}}$ is used as a conditional input to the conditional generative network $G_A(\cdot)$ to generate the reconstructed semantic signal, which can be expressed as 
\begin{equation}
    {\mathbf {\widetilde{y}}} = G_A({\mathbf{\hat{y}}}).
\end{equation}
The mean squared error (MSE) loss is used to train the proposed conditional generative network, 
which can be expressed as 
\begin{equation}
    \mathcal{L}_{\text{MSE}}  = \frac{1}{N}\sum\limits_{i = 1}^N \bigg\| \mathbf{y}_i - \mathbf{{\widetilde{y}}}_i \bigg\| _2^2, 
\end{equation}
where $N$ is the sample size, $\mathbf{y}_i$ is the $i$-th transmitted sample and $\mathbf{{\widetilde{y}}}_i$ is the $i$-th recovered sample. The Adam optimization algorithm is also used to iteratively update the parameters of $G_A(\cdot)$.
Similar to the training of $D_A(\cdot)$, the training of semantic error detector $D_{\bm{\delta}}({\bf \cdot})$ also adopt the loss function $\mathcal{L}_{\text{BCE}}$ to update parameters. A mini batch of sentence is selected from $\mathcal{K}_3$ and the observed data $\bf \hat{S}$ is entered into $D_{\bm{\delta}}({\bf \cdot})$ to obtain the confidence score $\hat{p}$ of ACK, which can be expressed as 
\begin{equation}
    \hat{p} = D_{\bm{\delta}}({\bf \hat{S}}).
\end{equation}
Apply the output $\hat{p}$ to \ref{loss} as the loss function of $D_{\bm{\delta}}({\bf \cdot})$ to optimize its parameters.

To ensure better performance of the trained module in practical applications, it is necessary to analyze its generalization error. 
Let $\mathcal{X}$ and $\mathcal{Y}$ denote the feature space and the label space, respectively, and let the hypothesis space be defined as $\mathcal{H} = \{h: \mathcal{X} \to \mathcal{Y}\}$.  The complexity of the hypothesis space can be assessed using the Vapnik–Chervonenkis (VC) dimension theory, a fundamental concept for analyzing generalization errors.
VC dimension theory provides a powerful framework for analyzing generalization errors, as stated in the following theorem \cite{bartlett2003vapnik}, 

\begin{theorem}\label{VC}
    Suppose the VC dimension of the hypothesis space $\mathcal{H}$ is finite and given by $d$, $m$ is the sample size, and let $h \in \mathcal{H}$. For any $m > d$ and $0 < \xi < 1$, we have
    \begin{equation}
        P\left(\left|E(h)-\widehat{E}(h)\right|\leqslant\sqrt{\frac{8d\ln\frac{2em}{d}+8\ln\frac{4}{\xi}}{m}}\right)\geqslant1-\xi,
    \end{equation}
    where $E(h)$ and $\widehat{E}(h)$ represent the semantic distortion generalization error and the semantic distortion empirical error, respectively.
\end{theorem}
According to Theorem \ref{VC}, we can conclude that $E(h)\leqslant\widehat{E}(h)+O\left(\sqrt{\frac{\ln(m/d)}{m/d}}\right)$ holds with probability of at least $1-\xi$, and furthermore, the rate of convergence of our algorithm is $O\left(\sqrt{\frac{\ln(m/d)}{m/d}}\right)$. More specifically, suppose the hypothesis space of our model is with finite VC dimension, the size of the VC dimension determines the convergence rate of the generalization error, i.e., the larger the VC dimension, the more complex the hypothesis space, and the slower the convergence rate of the generalization error.

\section{Simulation Results}\label{sec:results}
\subsection{Settings and Dataset}
In simulations, we assume that the ITS system includes a transmitter and a receiver, and all links undergo Rayleigh fading channels with $\mathbb E\{|h|^2\}=1$. We employ the dataset of proceedings of the European Parliament for training and testing \cite{koehn2005europarl}. The dataset comprises approximately 2 million sentences, each with 4 to 30 words. The dataset is split into training data and testing data, wherein the testing data contains around 60,000 sentences and the rest of the sentences are used for the training. Moreover, the semantic encoder and decoder follow a Transformer-based architecture, consisting of three encoding and decoding layers, each equipped with eight attention heads and dense layers. We refer to \cite{xie2021deep} for the parameter settings used in our experiments. More specifically, the network and training parameters are set as $L=30$, $\mathcal{V}=128$, $D=16$, and the learning rate is $\eta=10^{-4}$. Besides, 1-gram is frequently utilized for calculating BLEU. In this paper, the weights of 1-gram are set as $u_1 = 1$ and $u_n = 0$ for $n \ne 1$.

\subsection{Training of Semantic Communication System}

The training procedure is described as follows. In each training epoch $i\in \left\{1,2,...\right\}$, all sentences in $\mathcal{S}$ are transmitted in batches. The transmitter samples sentences from $\mathcal{S}$ to form a training batch $\mathcal{S}_i$. Before semantic encoding, each sentence $\mathbf{S} \in \mathcal{S}_i$ is tokenized into words, referred to as tokens. Each token is indexed by the vocabulary and has a unique integer index $\mathbf{w} \in {\mathbb{W}}$, where ${\mathbb{W}} = \left\{1,2,\dots, \mathcal{W}\right\} \subseteq \mathbb{Z}^+$, with $\mathcal{W}$ being the vocabulary size and $\mathbb{Z}^+$ denoting the set of positive integers. In other words, the sentence can be represented as $\mathbf{S} = [\mathbf{w}_1, \mathbf{w}_2, \dots, \mathbf{w}_L]$. To facilitate processing, each sentence within the same batch is padded to the maximum length $L$ with special symbols. 
Through this process, natural language is transformed into elements within the mathematical space ${\mathbb{W}}$. Simply put, any token in the vocabulary can be represented as a one-hot vector of length $\mathcal{W}$. Consequently, a sentence $\mathbf{S}$ can be expressed as a matrix $\mathbf{M} \in \mathbb{R}^{L \times \mathcal{W}}$. Since $\mathcal{W}$ is typically very large, $\mathbf{M}$ is inherently a sparse matrix. To address this issue, the word embedding method is employed in place of the one-hot vector, mapping each token to a dense vector of size $D$, where $D \ll \mathcal{W}$. The resulting embedded sentence matrix $\mathbf{E} \in \mathbb{R}^{L \times D}$ is denoted as $\mathbf{E} = Embed(\mathbf{S})$. 
The transmitter then encodes the embedded sentence $\mathbf{E}$ as $\mathbf{x} = f_{\bm{\alpha}}({\mathbf{E}})$, and the encoded semantic signal $\mathbf{x}$ is transmitted. Upon receiving the semantic signal $\mathbf{y}$, which has been affected by the communication channel, the receiver decodes it as $\mathbf{\hat{S}} = g_{\bm{\beta}}({\mathbf{y}})$. The receiver's primary objective is text reconstruction, aiming to recover the original sentence while minimizing semantic information loss. To enable joint training and optimization of the parameters for the semantic encoder and semantic decoder, the cross-entropy loss is also adopted, which can be formulated as 
\begin{align}
    \mathcal{L}_{\text{CE}}({\bf S}, \mathbf{\hat{S}}) = &-\sum_{l} \bigg( p(\mathbf{w}_l)\log q(\mathbf{w}_l) \notag \\
    &+ (1-p(\mathbf{w}_l))\log (1-q(\mathbf{w}_l))\bigg)  ,
\end{align}
where $\mathbf{w}_l$ is the $l$-th word, $p(\mathbf{w}_l)$ is its true probability and $q(\mathbf{w}_l)$ is the predicted probability.
Besides, we use the Adam optimization algorithm to iteratively update the parameters during the training.

\subsection{Verifications of Sem-HARQ Schemes}
\begin{figure}[!htb]
    \centering
    \includegraphics[width=3.0in]{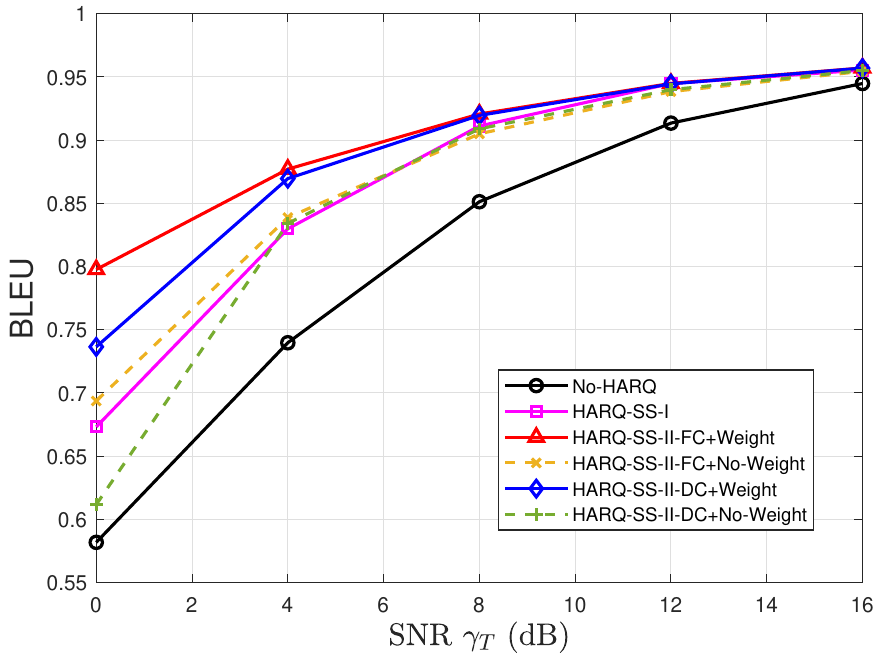}
    \caption{BLEU versus SNR for different HARQ schemes.}
    \label{BLEU}
\end{figure}

\begin{figure}[!htb]
    \centering
    \includegraphics[width=3.0in]{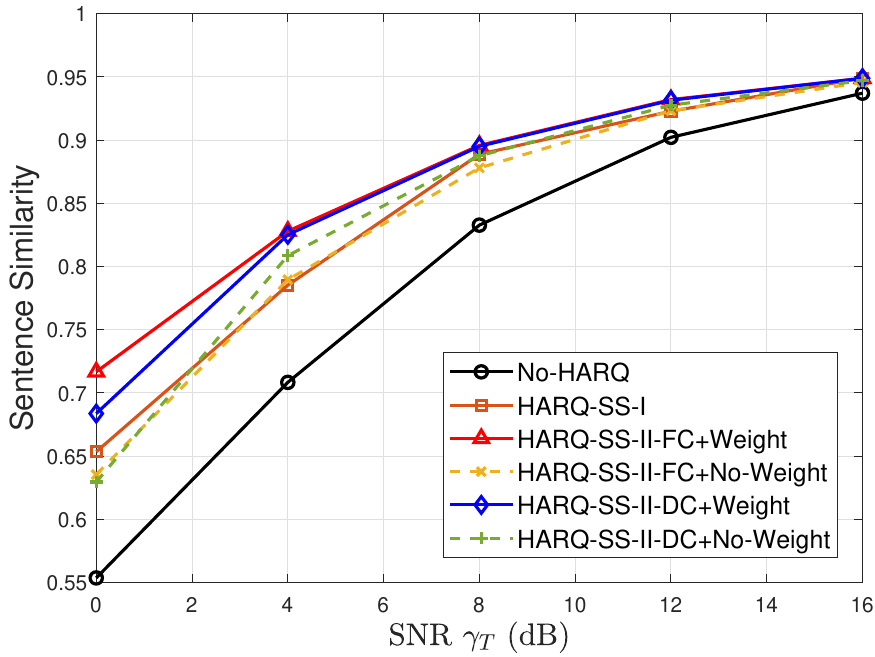}
    \caption{Sentence similarity versus SNR for different HARQ schemes.}
    \label{S_S}
\end{figure}

Figs. \ref{BLEU} and \ref{S_S} plot the BLEU scores and sentence similarity versus the SNR $\gamma_T \triangleq P/\mathcal N$. For benchmarking
comparison, the JSCC-enabled semantic communication system without HARQ is adopted as a baseline (labeled as ``No-HARQ''). This baseline employs the same semantic encoder and semantic decoder without a semantic error detector and channel data adaptation. All other approaches employ the proposed sem-HARQ system with proposed sem-HARQ-I and sem-HARQ-WC. The schemes of sem-HARQ-WC apply different combining approaches, with the confidence score weighted combination is labeled as ``weight'' and the equal combination method is labeled as ``No-weight''. As shown in Fig. \ref{BLEU}, all sem-HARQ-based approaches outperform the No-HARQ approaches in terms of BLEU score, demonstrating the effectiveness of the proposed sem-HARQ approach. Specifically, the sem-HARQ-based approaches show greater performance improvements at low SNR, with the sem-HARQ-WC-FC with Weight method achieving a $37.8\%$ improvement at an SNR of $0$dB. At high SNR, the performance of all approaches converges, as the improved channel conditions essentially eliminate the need for retransmissions. Comparing sem-HARQ-I and sem-HARQ-WC, it is clear that the sem-HARQ-WC scheme outperforms the former. This is because it utilizes prior information, albeit at the expense of higher memory requirements, which aligns with our expectations. In particular, at 0 dB SNR, the sem-HARQ-WC-FC with equal combining performs worse than the sem-HARQ-I scheme. This suggests that at low SNR, the lack of confidence score weighted combining results in information recovery being negatively impacted by erroneous semantics. Additionally, when comparing the two levels of merging, feature-level merging consistently outperforms decision-level merging, irrespective of the combination method used. This indicates that the semantic decoder benefits more from reinterpreting the merged semantic features, whereas decoding first limits this potential. Furthermore, when comparing the two combination methods, it is evident that the proposed confidence score weighted combination outperforms the equal combination, although the former entails a more complex process and longer decoding times. In fig. \ref{S_S}, the trend in sentence similarity aligns closely with that of the BLEU scores. Once again, all HARQ schemes outperform the No-HARQ scheme, with the best achieving a $29.5\%$ improvement in sentence similarity at 0 dB SNR. However, it can be observed that the gap in sentence similarity between the schemes narrows. This indicates that the sem-HARQ scheme focuses on achieving convergence in sentence-level enhancements, even if variations exist in detailed syntactic reconstruction. Overall, consistent with the BLEU scores, feature-level merging continues to outperform decision-level merging, while confidence score weighted combining demonstrates superior performance compared to equal combining. This effect is particularly pronounced at low SNR. The convergence of sentence similarity for all schemes in the high SNR scenario aligns with the convergence observed in the BLEU scores. However, the sem-HARQ schemes achieve values exceeding 0.9 around 8dB, both in terms of BLEU scores and sentence similarity, while the No-HARQ scheme requires 12dB. This further highlights the effectiveness and competitiveness of the proposed sem-HARQ schemes.

\begin{figure}[!htb]
    \centering
    \includegraphics[width=3.0in]{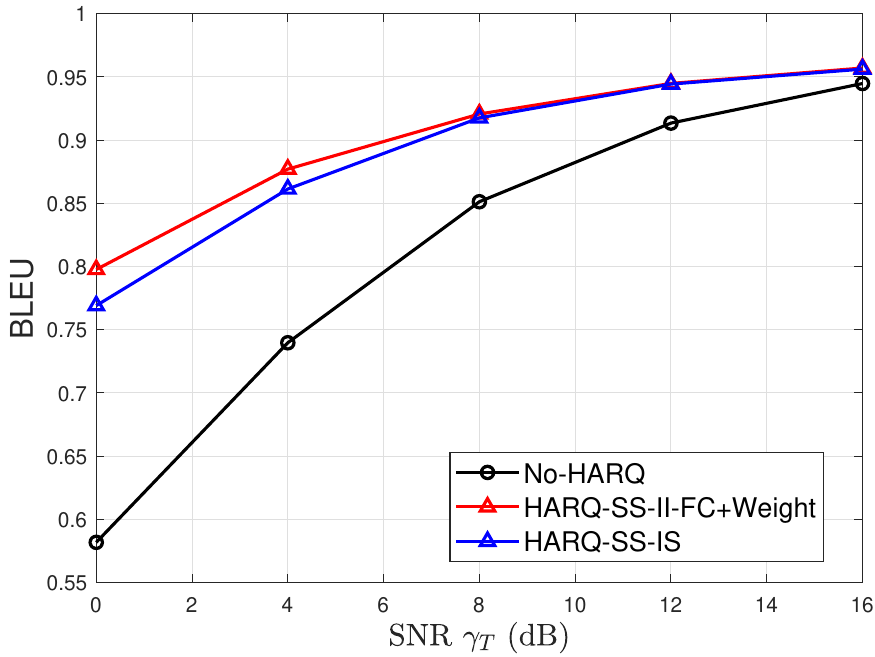}
    \caption{BLEU versus SNR for sem-HARQ-SC.}
    \label{synBLEU}
\end{figure}

\begin{figure}[!htb]
    \centering
    \includegraphics[width=3.0in]{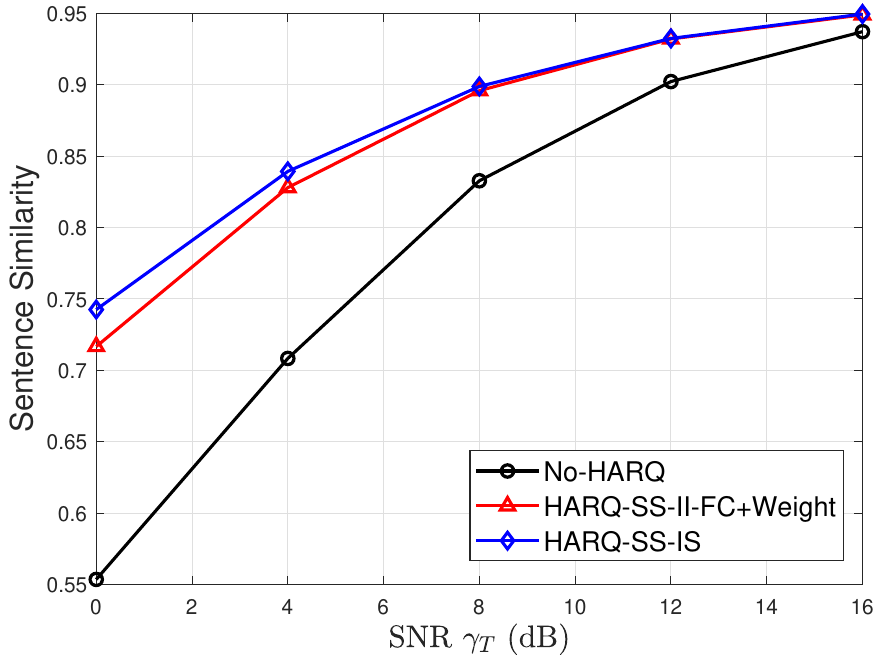}
    \caption{Sentence similarity versus SNR for sem-HARQ-SC.}
    \label{synS_S}
\end{figure}

Fig. \ref{synBLEU} and \ref{synS_S} illustrate the performance of sem-HARQ-SC on BLEU and sentence similarity versus the SNR, respectively. sem-HARQ-SC is fundamentally distinct from the previously described scheme. Specifically, each retransmission represents a different version of the semantics. In this paper, we proposed an adaptive synonym replacement approach, which enhances sentence recovery by progressively increasing synonym redundancy. Moreover, sem-HARQ-SC performs feature-level merging and employs a weighted combining method. To better demonstrate its performance, it is compared with the baseline ``No-HARQ'' and the previously best-performing sem-HARQ-WC-FC with the Weight method. It is evident that the sem-HARQ-SC scheme shows significant improvement over ``No-HARQ'', particularly at low SNR. At 0 dB SNR, the BLEU score shows an improvement of 32.2\%, while sentence similarity increases by 34.1\%, highlighting the significant enhancement. Notably, sem-HARQ-SC performs worse than sem-HARQ-WC-FC with the Weight method in terms of BLEU scores, while achieving better performance in sentence similarity. This indicates that the introduction of synonymous redundancy makes the recovered sentences more prone to synonymous substitutions, leading to a gain in preserving the overall meaning of the sentences. 



\subsection{Ablation Experiment}
\begin{figure}[!htb]
    \centering
    \includegraphics[width=3.0in]{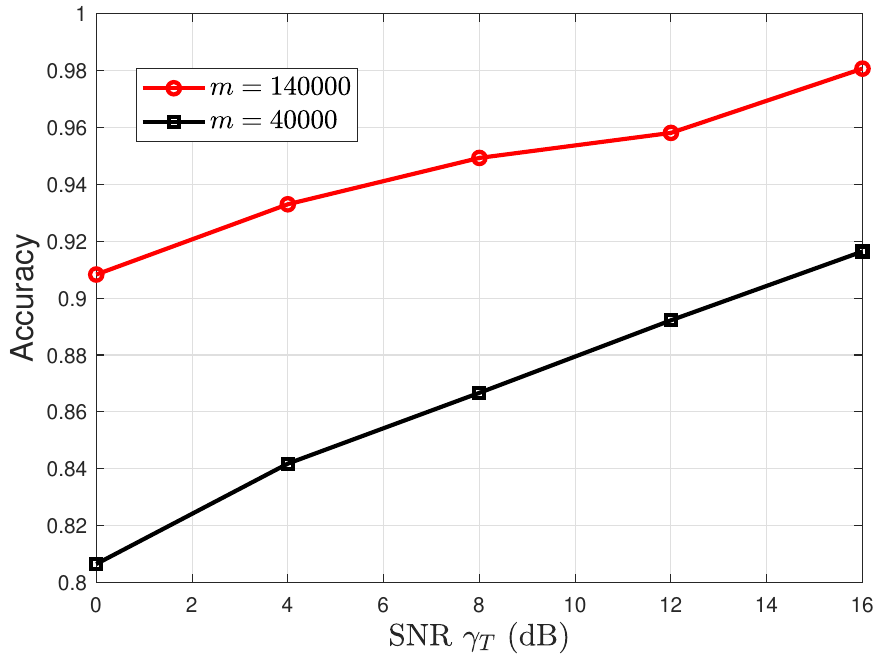}
    \caption{Accuracy versus SNR for different sample numbers.}
    \label{ACC}
\end{figure}

\begin{figure}[!htb]
    \centering
    \includegraphics[width=3.0in]{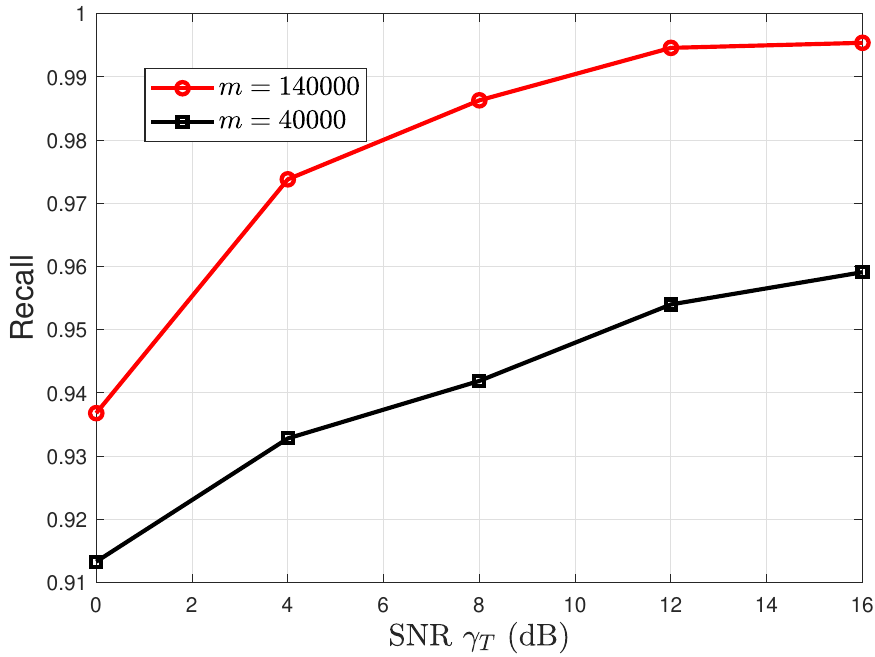}
    \caption{Recall versus SNR for different sample numbers.}
    \label{REC}
\end{figure}

To evaluate the performance of the proposed semantic error detector, common metrics used to assess classification network performance were employed, accuracy and recall. Accuracy measures the proportion of correctly predicted samples out of the total samples, providing an overall assessment of the model's correctness.
The indicator can be calculated as
\begin{equation}
    \operatorname{Accuracy} = \frac{TP+TN}{TP+TN+FP+FN} 
\end{equation}
where $TP$ is the true positive, i.e., sentence semantic correct with feedback ACK, $TN$ is the true negative, i.e., sentence semantic error with feedback NACK, $FP$ is the false positive, i.e., sentence semantic error with feedback ACK and $FN$ is the false negative, i.e., sentence semantic correct with feedback NACK. Fig. \ref{ACC} illustrates the accuracy versus SNR for different sample sizes $m$. As observed, accuracy improves consistently with an increase in sample size, aligning with the theoretical analysis presented earlier. When the sample size reaches fourteen million, the accuracy exceeds 0.9, demonstrating that the semantic error detector effectively identifies whether the received sentence's semantics are correct. It is important to note that accuracy improves as SNR increases. This phenomenon can be attributed to the higher probability of the received sentence being semantically correct at higher SNR levels. As a result, collecting semantically incorrect samples becomes more challenging at high SNR. This may result in sample imbalance, causing the semantic error detector to be more inclined to classify the received sentence as correct, which consequently leads to a higher accuracy rate.
To better evaluate the performance in cases of sample imbalance, Recall, also known as sensitivity, was adopted as an additional indicator. Recall represents the proportion of semantically correct sentences that are correctly identified, and can be calculated as
\begin{equation}
    \operatorname{Recall} = \frac{TP}{TP+FN}.
\end{equation}
Recall assesses the model's ability to correctly identify positive instances. A higher recall value indicates that the model is more effective in detecting positive samples. Fig. \ref{REC} illustrates the Recall as a function of SNR for different sample sizes. As expected, an increase in sample size leads to a higher Recall, which further supports our analysis. It is noticeable that the Recall values are consistently above 0.9, which surpasses the accuracy. This indicates that the semantic error detector is more effective in identifying positive instances, which helps explain the observed increase in accuracy with higher SNR.

\begin{figure}[!htb]
    \centering
    \includegraphics[width=3.0in]{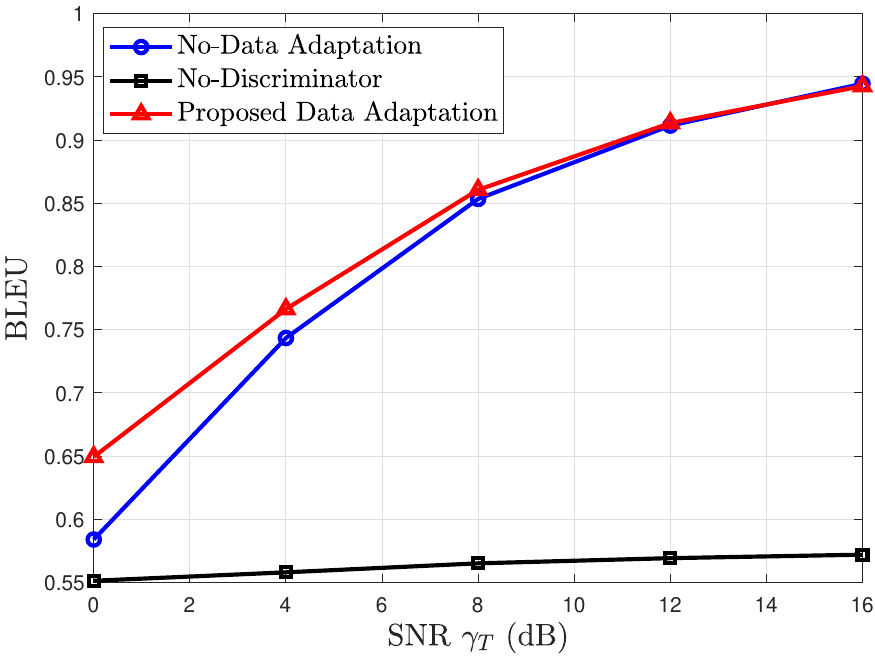}
    \caption{BLEU score versus SNR for different variants of the sem-HARQ system.}
    \label{ADA}
\end{figure}

Fig. \ref{ADA} illustrates the relationship between BLEU scores versus SNR for different variants of the sem-HARQ system. The three configurations include the sem-HARQ system with complete Data Adaptation (labeled as ``With-Data Adaptation''), the system without the Data Adaptation module (labeled as ``No-Data Adaptation''), and the system with the Data Adaptation that only corrector (labeled as ``No-Discriminator''). To provide a clearer visualization of the impact of Data Adaptation, none of the variants involve retransmissions. It is clear that the sem-HARQ system with complete Data Adaptation outperforms all other variants across all SNR, particularly at low SNR. At high SNR, the BLEU scores of the sem-HARQ system with complete Data Adaptation converge with those of the system without Data Adaptation. This can be attributed to the fact that at high SNR, improved channel conditions result in fewer semantic errors, thereby diminishing the contribution of Data Adaptation. It is worth noting that when only the corrector is used in Data Adaptation, meaning the received semantics are directly fed into the corrector without any discrimination, the BLEU score is lower compared to the system without Data Adaptation. This suggests that while the corrector effectively corrects incorrect semantics, feeding all semantics into it without any filtering can inadvertently lead to the conversion of correct semantics into incorrect ones. This phenomenon is more pronounced at high SNR, consistent with the earlier analysis. At high SNR, the probability of correctly interpreting semantics increases, leading to a higher likelihood of such misconversions.

\section{Conclusion}\label{sec:con}

In this paper, to address the reliable communication challenges in ITS, we propose a JSCC-enabled sem-HARQ system designed to enhance reliable communication. We introduce a novel conditional generative model for signal reconstruction, enabling semantic-aware error detection and correction. Within this framework, we design three distinct semantic HARQ schemes: sem-HARQ-I, sem-HARQ-WC, and sem-HARQ-SC. Specifically, Sem-HARQ-I employs identical semantic retransmissions while discarding erroneous semantics in each round to minimize memory usage. In contrast, sem-HARQ-WC retains erroneous semantics across retransmissions to improve decoding accuracy, while sem-HARQ-SC conveys different semantic representations in each retransmission by replacing words with synonyms or rephrasing sentences, thereby reducing ambiguity and improving overall transmission performance.

To further enhance reliability, we develop a semantic error detector and a reconstructor module, both leveraging a constructed local knowledge base. The semantic error detector identifies errors at the semantic level, departing from traditional bit-level error detection, and generates feedback messages to inform the need for retransmissions. The reconstructor module enables partial correction of semantic errors, reducing the necessity for retransmissions and improving efficiency. Simulation results validate the effectiveness of the proposed sem-HARQ system, demonstrating significant performance improvements across various scenarios. By integrating semantically-aware error correction and retransmission strategies, sem-HARQ can significantly enhances ITS by reducing communication overhead, improving robustness in noisy environments, enabling more effective data fusion, and ensuring scalable, reliable V2X communications.

\bibliographystyle{IEEEtran}
 	\bibliography{ref}

\end{document}